\documentclass[12pt]{article}

\usepackage{cite}
\usepackage{epsfig}
\usepackage{amsmath}
\usepackage{eufrak}
\usepackage{pstricks}
\usepackage{afterpage}

\newcommand{\mycaption}[1]{\caption{\sl #1}}

\makeatletter
\def\section{\@startsection {section}{1}{\z@}{+3.0ex plus +1ex minus
  +.2ex}{2.3ex plus .2ex}{\large\bf\boldmath}}
\def\subsection{\@startsection{subsection}{2}{\z@}{+2.5ex plus +1ex
minus +.2ex}{1.5ex plus .2ex}{\normalsize\bf\boldmath}}
\def\subsubsection{\@startsection{subsubsection}{3}{\z@}{+3.25ex plus
 +1ex minus +.2ex}{1.5ex plus .2ex}{\normalsize\it}}

\makeatletter
\newcommand\footnoteref[1]{\protected@xdef\@thefnmark{\ref{#1}}\@footnotemark}
\makeatother

\graphicspath{{.}{plots/}}

\begin{document}
\thispagestyle{empty}

\def\thefootnote{\fnsymbol{footnote}}

\begin{flushright}
\end{flushright}

\vspace{1cm}

\begin{center}

{\large {\bf TVID 2: Evaluation of planar-type three-loop self-energy
integrals with arbitrary masses}}
\\[3.5em]
{\large
Stefan Bauberger$^1$, Ayres~Freitas$^{2,}$\footnote{Email: {\tt
afreitas@pitt.edu}}, Daniel Wiegand$^{3,4,}$\footnote{Email: {\tt
daniel.wiegand@northwestern.edu}}
}

\vspace*{1cm}

{\sl
$^1$ Hochschule f\"ur Philosophie,
Philosophische Fakult\"at S.J.,
Kaulbachstr.\ 31,\\ 80539 M\"unchen, Germany \\[1ex]
$^2$ Pittsburgh Particle-physics Astro-physics \& Cosmology Center
(PITT-PACC),\\ Department of Physics \& Astronomy, University of Pittsburgh,
Pittsburgh, PA 15260, USA\\[1ex]
$^3$ HEP Division, Argonne National Laboratory, Argonne, Illinois 60439, USA\\[1ex]
$^4$ Department of Physics \& Astronomy, Northwestern University,\\ Evanston, Illinois 60208, USA
}

\end{center}

\vspace*{2.5cm}

\begin{abstract}
We present TVID 2, a program to numerically evaluate an important class of
planar three-loop self-energy master integrals with arbitrary masses. As with
the predecessor version (TVID 1) the integrals are separated
into a known piece, containing the UV divergencies, and a finite
piece that is integrated numerically, implemented in C. The set of master
integrals under consideration was found with self-energy diagrams containing two
closed fermion loops in mind. Two techniques are employed in deriving
the expressions for the finite pieces that are then numerically integrated:
(a) Sub-loop dispersion relations in the case of topologies containing
sub-bubbles, and (b) a modification of the procedure suggested by Ghinculov 
for integrals with only sub-loop triangles.
\end{abstract}

\setcounter{page}{0}
\setcounter{footnote}{0}

\newpage

%%%%%%%%%%%%%%%%%%%%%%%%%%%%%%%%%%%%%%%%%%%%%%%%%%%%%%%%%%%%%%

\section{Introduction}

\noindent
The calculation of higher-order radiative corrections is important for the
interpretation of precision measurements at the LHC and various $e^+e^-$ machines, such
as SuperKEKB and planned future Higgs and $Z$ factories. Multi-loop
contributions in the full Standard Model or models beyond the Standard Model
(BSM) are particularly challenging due to the presence of many independent mass
and momentum scales \cite{review}. General loop integrals beyond the one-loop level cannot be solved
analytically in terms of elementary functions. This observation prompted the
investigation of new classes of special functions, such as harmonic
polylogarithms \cite{Remiddi:1999ew}, generalized harmonic
polylogarithms \cite{ghpl}, and elliptic polylogarithms \cite{epl}, see $e.\,g.$
Ref.~\cite{Blumlein:2019svg} for a recent review. However, it
is not clear if any multi-loop integral can be represented by these classes of
functions, in particular beyond the two-loop level.

This motivates the development of numerical methods for multi-loop integrations.
Two general approaches, which in principle can be applied to any number of loops
and external legs, are known: sector decomposition and Mellin-Barnes
representations. The former has been realized in the {\sc SecDec} \cite{secdec}
and FIESTA \cite{fiesta,fiesta4} software packages, while the latter is the
basis for the {\sc AMBRE/MBnumerics} project \cite{ambre}. Both approaches
provide an algorithmic procedure for removing UV and IR singularities, but they
require very significant computing resources, especially for integrals with
physical internal thresholds that develop imaginary parts. Alternatively, more
efficient numerical integration methods can be developed for limited classes of
multi-loop integrals (see Ref.~\cite{review} for a review of some of these
methods). At the three-loop level, an important step in this direction was
achieved with the programs TVID \cite{3lvac,Bauberger:2017nct} and 3VIL \cite{3vil}, which
can evaluate the master integrals for arbitrary three-loop vacuum integrals.

This article reports on the new version 2.0 of TVID,  which includes a large
class of three-loop self-energy master integrals. This class consists of the
master integrals necessary to evaluate three-loop self-energy diagrams
containing two closed fermion loops. They are descendants of the planar
three-loop self-energy topology ($i.\,e.$ the ladder topology). It is shown that
all these master integrals can be evaluated in terms of at most two-dimensional
numerical integrals, by making use of the following two ideas: 
\begin{itemize}
\item Integrals with sub-loop self-energies are
evaluated by using dispersion relations for the sub-loop. This technique was
previously developed for the evaluation of two-loop self-energy master integrals
\cite{disp2,disp2a}.
\item Planar integrals with without sub-loop self-energies are tackled with a
variant of the method introduced in Ref.~\cite{Ghinculov:1996vd}.
\end{itemize}
See section~\ref{sc:def} for more details on the master integrals that fall into
either of these two categories. The construction of the numerical integral
representations for these master integrals is illustrated for a few
characteristic examples in section~\ref{sc:ex}. It should be noted that many of
the master integrals are UV divergent, and these singularities must be removed
before the numerical integration can be carried out. In TVID, this is achieved
by subtracting terms that have the same singularity structure, but that lead to
simpler integrals which are already known in the literature, see
section~\ref{sc:ex} and appendix~\ref{sc:div}. TVID provides the integrated
subtraction terms in the framework of dimensional regularization and then
numerically evaluates the finite remainder integrals. More information on the
implementation of the three-loop self-energy integrals in TVID 2.0 can be found
in section \ref{sc:tvid}, together with a discussion of potential problems and
comparison to results in the literature. A manual for the installation and usage
of TVID 2.0 is provided in appendix~\ref{sc:manual}.

%%%%%%%%%%%%%%%%%%%%%%%%%%%%%%%%%%%%%%%%%%%%%%%%%%%%%%%%%%%%%%
\begin{figure}[t]
\begin{center}
\psset{linewidth=1pt}
\psset{dotsize=5pt}
\begin{tabular}{p{3.5cm}p{3.5cm}p{3.5cm}p{3.5cm}}
&&& \\[-.2cm]
\begin{center}
\rput[rb](-1.8,-0.1){$\stackrel{{\displaystyle p}}{\rightarrow}$}%
\pscircle(0,0){1}%
\psdot(1,0)%
\psdot(-1,0)%
\psdot(-1,0)%
\psline(-1.5,0)(-1,0)%
\psline(1.5,0)(1,0)%
\psarc(0,-1){1.414}{45}{135}%
\psarc(0,1){1.414}{-135}{-45}%
\rput[c](0,-0.75){\small 8}%
\rput[c](0,-0.2){\small 6}% 
\rput[c](0,0.65){\small 3}% 
\rput[c](0,1.2){\small 1}%
\end{center}
&
\begin{center}
\pscircle(0,0){1}%
\psdot(1,0)%
\psdot(-1,0)%
\psdot(0,-1)%
\psline(-1.5,0)(-1,0)%
\psline(1.5,0)(1,0)%
\psarc(-1,-1){1}{0}{90}%
\psarc(1,-1){1}{90}{180}%
\rput[r](-0.9,-0.6){\small 1}%
\rput[r](-0.35,-0.4){\small 3}% 
\rput[l](0.9,-0.6){\small 7}%
\rput[l](0.35,-0.4){\small 6}% 
\rput[t](0,0.9){\small 5}% 
\end{center}
&
\begin{center}
\pscircle(0,0){1}%
\psdot(1,0)%
\psdot(-1,0)%
\psdot(0,-1)%
\psline(-1.5,0)(1.5,0)%
\psarc(1,-1){1}{90}{180}%
\rput[r](-0.9,-0.6){\small 4}%
\rput[c](0,0.2){\small 3}% 
\rput[l](0.9,-0.6){\small 7}%
\rput[l](0.35,-0.4){\small 6}% 
\rput[t](0,0.9){\small 2}% 
\end{center}
&
\begin{center}
\pscircle(0,0){1}%
\psdot(1,0)%
\psdot(-1,0)%
\psdot(-0.5,-0.866)%
\psline(-1.5,0)(-1,0)%
\psline(1.5,0)(1,0)%
\pscurve(1,0)(0,0)(-0.5,-0.866)%
\psline(1,0)(-0.5,-0.866)%
\rput[r](-0.9,-0.6){\small 1}%
\rput[c](-0.2,0.2){\small 3}% 
\rput[l](0.9,-0.6){\small 7}%
\rput[r](0.2,-0.25){\small 6}% 
\rput[t](0,0.9){\small 2}% 
\end{center}
\\[0.6cm]
\centering $U_{\rm 4a}$ & 
\centering $U_{\rm 5a}$ & 
\centering $U_{\rm 5b}$ & 
\centering $U_{\rm 5c}$ 
\cr\\[0.3cm]
\begin{center}
\pscircle(0,0){1}%
\psdot(1,0)%
\psdot(-1,0)%
\psdot(0.35,-0.9)%
\psdot(-0.35,-0.9)%
\psline(-1.5,0)(-1,0)%
\psline(1.5,0)(1,0)%
\psarc(-1,-0.7){0.7}{-20}{90}%
\psarc(1,-0.7){0.7}{90}{200}%
\rput[r](-0.9,-0.6){\small 1}%
\rput[l](-0.45,0){\small 3}% 
\rput[l](0.9,-0.6){\small 7}%
\rput[r](0.45,0){\small 6}% 
\rput[t](0,0.9){\small 5}% 
\rput[b](0,-0.85){\small 4}% 
\end{center}
&
\begin{center}
\pscircle(0,0){1}%
\psdot(1,0)%
\psdot(-1,0)%
\psdot(0,-1)%
\psdot(0,1)%
\psline(-1.5,0)(-1,0)%
\psline(1.5,0)(1,0)%
\psarc(1,-1){1}{90}{180}%
\psarc(-1,1){1}{-90}{0}%
\rput[r](-0.9,-0.7){\small 5}%
\rput[l](0.9,0.7){\small 4}%
\rput[r](-0.35,0.45){\small 3}% 
\rput[l](0.9,-0.6){\small 8}%
\rput[l](0.35,-0.4){\small 6}% 
\rput[r](-0.9,0.65){\small 1}% 
\end{center}
&
\begin{center}
\pscircle(0,0){1}%
\psdot(1,0)%
\psdot(-1,0)%
\psdot(-0.5,-0.866)%
\psdot(0,-0.577)%
\psline(-1.5,0)(-1,0)%
\psline(1.5,0)(1,0)%
\pscurve(0,-0.577)(0.2,0.2)(1,0)%
\psline(1,0)(-0.5,-0.866)%
\rput[r](-0.9,-0.6){\small 1}%
\rput[r](-0.25,-0.45){\small 4}% 
\rput[l](0.9,-0.6){\small 3}%
\rput[b](-0.1,0){\small 6}% 
\rput[r](0.5,-0.1){\small 7}% 
\rput[t](0,0.9){\small 2}% 
\end{center}
&
\begin{center}
\pscircle(0,0){1}%
\psdot(1,0)%
\psdot(-1,0)%
\psdot(0,1)%
\psdot(0,-1)%
\psline(-1.5,0)(-1,0)%
\psline(1.5,0)(1,0)%
\psline(0,-1)(0,1)%
\psarc(1,-1){1}{90}{180}%
\rput[r](-0.9,0.7){\small 1}%
\rput[r](-0.9,-0.7){\small 2}%
\rput[rb](-0.05,0){\small 3}% 
\rput[l](0.9,-0.7){\small 8}%
\rput[l](0.35,-0.4){\small 6}% 
\rput[l](0.9,0.7){\small 4}% 
\end{center}
\\[0.6cm]
\centering $U_{\rm 6a}$ & 
\centering $U_{\rm 6b}$ & 
\centering $U_{\rm 6c}$ &
\centering $U_{\rm 6m}$
\cr\\[0.3cm]
\begin{center}
\pscircle(0,0){1}%
\psdot(1,0)%
\psdot(-1,0)%
\psdot(0,1)%
\psdot(0,-1)%
\psline(-1.5,0)(-1,0)%
\psline(1.5,0)(1,0)%
\psarc(-1,0){1.414}{-45}{45}%
\psarc(1,0){1.414}{135}{-135}%
\rput[r](-0.9,0.7){\small 1}%
\rput[r](-0.9,-0.7){\small 2}%
\rput[r](-0.5,0){\small 3}% 
\rput[r](0.3,0){\small 6}% 
\rput[l](0.9,0.7){\small 7}%
\rput[l](0.9,-0.7){\small 8}%
\end{center}
&
\begin{center}
\pscircle(0,0){1}%
\psdot(1,0)%
\psdot(-1,0)%
\psdot(-0.3,0.954)%
\psdot(-0.3,-0.954)%
\psdot(0.35,-0.9)%
\psline(-1.5,0)(-1,0)%
\psline(1.5,0)(1,0)%
\psline(-0.3,-0.954)(-0.3,0.954)%
\psarc(1,-0.7){0.7}{90}{200}%
\rput[r](-0.9,0.7){\small 1}%
\rput[r](-0.9,-0.7){\small 2}%
\rput[lb](-0.25,0){\small 3}% 
\rput[b](0,-0.85){\small 5}% 
\rput[l](0.9,-0.7){\small 8}%
\rput[r](0.45,0){\small 6}% 
\rput[l](0.9,0.7){\small 4}% 
\end{center}
&
\begin{center}
\pscircle(0,0){1}%
\psdot(1,0)%
\psdot(-1,0)%
\psdot(-0.5,0.866)%
\psdot(0,-1)%
\psdot(0.5,0.866)%
\psline(-1.5,0)(-1,0)%
\psline(1.5,0)(1,0)%
\psline(0,-1)(-0.5,0.866)%
\psline(0,-1)(0.5,0.866)%
\rput[r](-0.9,0.7){\small 1}%
\rput[r](-0.9,-0.7){\small 2}%
\rput[r](-0.35,0){\small 3}% 
\rput[t](0,0.85){\small 4}% 
\rput[l](0.35,0){\small 6}% 
\rput[l](0.9,0.7){\small 7}% 
\rput[l](0.9,-0.7){\small 8}%
\end{center}
&
\begin{center}
\pscircle(0,0){1}%
\psdot(1,0)%
\psdot(-1,0)%
\psdot(-0.3,0.954)%
\psdot(-0.3,-0.954)%
\psdot(0.3,0.954)%
\psdot(0.3,-0.954)%
\psline(-1.5,0)(-1,0)%
\psline(1.5,0)(1,0)%
\psline(-0.3,-0.954)(-0.3,0.954)%
\psline(0.3,-0.954)(0.3,0.954)%
\rput[r](-0.9,0.7){\small 1}%
\rput[r](-0.9,-0.7){\small 2}%
\rput[r](-0.35,0){\small 3}% 
\rput[t](0,0.85){\small 4}% 
\rput[b](0,-0.85){\small 5}% 
\rput[l](0.35,0){\small 6}% 
\rput[l](0.9,0.7){\small 7}% 
\rput[l](0.9,-0.7){\small 8}%
\end{center}
\\[0.6cm]
\centering $U_{\rm 6n}$ &
\centering $U_{\rm 7m}$ &
\centering $U_{\rm 7a}$ &
\centering $U_{\rm 8a}$
\end{tabular}
\end{center}
\vspace{-2.5ex}
\mycaption{Basic master integral topologies without doubled propagators considered in this paper. 
\label{fig:diag1}}
\end{figure}
%%%%%%%%%%%%%%%%%%%%%%%%%%%%%%%%%%%%%%%%%%%%%%%%%%%%%%%%%%%%%%
\begin{figure}[t]
\begin{center}
\psset{linewidth=1pt}
\psset{dotsize=5pt}
\begin{tabular}{p{3.5cm}p{3.5cm}p{3.5cm}p{3.5cm}}
&&& \\[-.2cm]
\begin{center}
\pscircle(0,0){1}%
\psdot(1,0)%
\psdot(-1,0)%
\psdot(-1,0)%
\psdot(0,-1)%
\psline(-1.5,0)(-1,0)%
\psline(1.5,0)(1,0)%
\psarc(0,-1){1.414}{45}{135}%
\psarc(0,1){1.414}{-135}{-45}%
\rput[r](-0.9,-0.7){\small 1}%
\rput[l](0.9,-0.7){\small 1}%
\rput[c](0,-0.2){\small 3}% 
\rput[c](0,0.65){\small 6}% 
\rput[c](0,1.2){\small 8}%
\end{center}
&
\begin{center}
\pscircle(0,0){1}%
\psdot(1,0)%
\psdot(-1,0)%
\psdot(-1,0)%
\psdot(0,-1)%
\psdot(0,-0.414)%
\psline(-1.5,0)(-1,0)%
\psline(1.5,0)(1,0)%
\psarc(0,-1){1.414}{45}{135}%
\psarc(0,1){1.414}{-135}{-45}%
\rput[r](-0.9,-0.7){\small 1}%
\rput[l](0.9,-0.7){\small 1}%
\rput[c](-0.3,-0.1){\small 3}% 
\rput[c](0.3,-0.1){\small 3}% 
\rput[c](0,0.65){\small 6}% 
\rput[c](0,1.2){\small 8}%
\end{center}
&
\begin{center}
\pscircle(0,0){1}%
\psdot(1,0)%
\psdot(-1,0)%
\psdot(-1,0)%
\psdot(-0.5,-0.866)%
\psdot(0.5,-0.866)%
\psline(-1.5,0)(-1,0)%
\psline(1.5,0)(1,0)%
\psarc(0,-1){1.414}{45}{135}%
\psarc(0,1){1.414}{-135}{-45}%
\rput[r](-0.9,-0.7){\small 1}%
\rput[l](0.9,-0.7){\small 1}%
\rput[c](0,-0.75){\small 1}%
\rput[c](0,-0.2){\small 3}% 
\rput[c](0,0.65){\small 6}% 
\rput[c](0,1.2){\small 8}%
\end{center}
&
\begin{center}
\pscircle(0,0){1}%
\psdot(1,0)%
\psdot(-1,0)%
\psdot(0,-1)%
\psdot(0,1)%
\psline(-1.5,0)(-1,0)%
\psline(1.5,0)(1,0)%
\psarc(-1,-1){1}{0}{90}%
\psarc(1,-1){1}{90}{180}%
\rput[r](-0.9,-0.6){\small 3}%
\rput[r](-0.35,-0.4){\small 1}% 
\rput[l](0.9,-0.6){\small 7}%
\rput[l](0.35,-0.4){\small 6}% 
\rput[r](-0.9,0.7){\small 5}%
\rput[l](0.9,0.7){\small 5}%
\end{center}
\\[0.6cm]
\centering $U_{\rm 4a1}$ &
\centering $U_{\rm 4a2}$ &
\centering $U_{\rm 4a3}$ &
\centering $U_{\rm 5a1}$
\cr\\[0.3cm]
\begin{center}
\pscircle(0,0){1}%
\psdot(1,0)%
\psdot(-1,0)%
\psdot(0,-1)%
\psdot(-0.294,-0.294)%
\psline(-1.5,0)(-1,0)%
\psline(1.5,0)(1,0)%
\psarc(-1,-1){1}{0}{90}%
\psarc(1,-1){1}{90}{180}%
\rput[r](-0.9,-0.6){\small 3}%
\rput[r](-0.15,-0.65){\small 1}% 
\rput[r](-0.35,0.1){\small 1}% 
\rput[l](0.9,-0.6){\small 7}%
\rput[l](0.35,-0.4){\small 6}% 
\rput[t](0,0.9){\small 5}% 
\end{center}
&
\begin{center}
\pscircle(0,0){1}%
\psdot(1,0)%
\psdot(-1,0)%
\psdot(0,-1)%
\psdot(-0.707,-0.707)%
\psline(-1.5,0)(1.5,0)%
\psarc(1,-1){1}{90}{180}%
\rput[l](-0.8,-0.3){\small 4}%
\rput[b](-0.3,-0.8){\small 4}%
\rput[c](0,0.2){\small 3}% 
\rput[l](0.9,-0.6){\small 7}%
\rput[l](0.35,-0.4){\small 6}% 
\rput[t](0,0.9){\small 2}% 
\end{center}
&
\begin{center}
\pscircle(0,0){1}%
\psdot(1,0)%
\psdot(-1,0)%
\psdot(0,-1)%
\psdot(0,1)%
\psline(-1.5,0)(1.5,0)%
\psarc(1,-1){1}{90}{180}%
\rput[l](-0.7,-0.5){\small 4}%
\rput[c](0,0.2){\small 3}% 
\rput[l](0.9,-0.6){\small 7}%
\rput[l](0.35,-0.4){\small 6}% 
\rput[r](-0.9,0.7){\small 2}%
\rput[l](0.9,0.7){\small 2}%
\end{center}
&
\begin{center}
\pscircle(0,0){1}%
\psdot(1,0)%
\psdot(-1,0)%
\psdot(-0.5,-0.866)%
\psdot(0.5,-0.866)%
\psline(-1.5,0)(-1,0)%
\psline(1.5,0)(1,0)%
\pscurve(1,0)(0,0)(-0.5,-0.866)%
\psline(1,0)(-0.5,-0.866)%
\rput[r](-0.9,-0.6){\small 1}%
\rput[c](-0.2,0.2){\small 3}% 
\rput[l](0.9,-0.6){\small 7}%
\rput[b](0.2,-0.85){\small 7}%
\rput[r](0.2,-0.25){\small 6}% 
\rput[t](0,0.9){\small 2}% 
\end{center}
\\[0.6cm]
\centering $U_{\rm 5a2}$ &
\centering $U_{\rm 5b1}$ &
\centering $U_{\rm 5b2}$ &
\centering $U_{\rm 5c1}$
\cr\\[0.3cm]
\begin{center}
\pscircle(0,0){1}%
\psdot(1,0)%
\psdot(-1,0)%
\psdot(0,1)%
\psdot(0,-1)%
\psdot(0,0)%
\psline(-1.5,0)(-1,0)%
\psline(1.5,0)(1,0)%
\psline(0,-1)(0,1)%
\psarc(1,-1){1}{90}{180}%
\rput[r](-0.9,0.7){\small 1}%
\rput[r](-0.9,-0.7){\small 2}%
\rput[r](-0.05,0.5){\small 3}% 
\rput[r](-0.05,-0.5){\small 3}% 
\rput[l](0.9,-0.7){\small 8}%
\rput[l](0.35,-0.4){\small 6}% 
\rput[l](0.9,0.7){\small 4}% 
\end{center}
&
\begin{center}
\pscircle(0,0){1}%
\psdot(1,0)%
\psdot(-1,0)%
\psdot(0,1)%
\psdot(0,-1)%
\psdot(0.294,-0.294)%
\psline(-1.5,0)(-1,0)%
\psline(1.5,0)(1,0)%
\psline(0,-1)(0,1)%
\psarc(1,-1){1}{90}{180}%
\rput[r](-0.9,0.7){\small 1}%
\rput[r](-0.9,-0.7){\small 2}%
\rput[rb](-0.05,0){\small 3}% 
\rput[l](0.9,-0.7){\small 8}%
\rput[l](0.15,-0.65){\small 6}% 
\rput[l](0.35,0.1){\small 6}% 
\rput[l](0.9,0.7){\small 4}% 
\end{center}
&
\begin{center}
\pscircle(0,0){1}%
\psdot(1,0)%
\psdot(-1,0)%
\psdot(0,1)%
\psdot(0,-1)%
\psdot(-0.707,-0.707)%
\psline(-1.5,0)(-1,0)%
\psline(1.5,0)(1,0)%
\psline(0,-1)(0,1)%
\psarc(1,-1){1}{90}{180}%
\rput[r](-0.9,0.7){\small 1}%
\rput[l](-0.8,-0.3){\small 2}%
\rput[b](-0.3,-0.8){\small 2}%
\rput[rb](-0.05,0){\small 3}% 
\rput[l](0.9,-0.7){\small 8}%
\rput[l](0.35,-0.4){\small 6}% 
\rput[l](0.9,0.7){\small 4}% 
\end{center}
&
\begin{center}
\pscircle(0,0){1}%
\psdot(1,0)%
\psdot(-1,0)%
\psdot(0,1)%
\psdot(0,-1)%
\psdot(-0.414,0)%
\psline(-1.5,0)(-1,0)%
\psline(1.5,0)(1,0)%
\psarc(-1,0){1.414}{-45}{45}%
\psarc(1,0){1.414}{135}{-135}%
\rput[r](-0.9,0.7){\small 1}%
\rput[r](-0.9,-0.7){\small 2}%
\rput[l](-0.25,0.4){\small 3}% 
\rput[l](-0.25,-0.4){\small 3}% 
\rput[l](0.5,0){\small 6}% 
\rput[l](0.9,0.7){\small 7}%
\rput[l](0.9,-0.7){\small 8}%
\end{center}
\\[0.6cm]
\centering $U_{\rm 6m1}$ & 
\centering $U_{\rm 6m2}$ & 
\centering $U_{\rm 6m3}$ &
\centering $U_{\rm 6n1}$
\cr\\[0.3em]
\begin{center}
\pscircle(0,0){1}%
\psdot(1,0)%
\psdot(-1,0)%
\psdot(0,1)%
\psdot(0,-1)%
\psdot(-0.707,-0.707)%
\psline(-1.5,0)(-1,0)%
\psline(1.5,0)(1,0)%
\psarc(-1,0){1.414}{-45}{45}%
\psarc(1,0){1.414}{135}{-135}%
\rput[r](-0.9,0.7){\small 1}%
\rput[r](-0.95,-0.45){\small 2}%
\rput[t](-0.5,-0.9){\small 2}%
\rput[l](-0.3,0){\small 3}% 
\rput[l](0.5,0){\small 6}% 
\rput[l](0.9,0.7){\small 7}%
\rput[l](0.9,-0.7){\small 8}%
\end{center}
&
\begin{center}
\pscircle(0,0){1}%
\psdot(1,0)%
\psdot(-1,0)%
\psdot(-0.5,0.866)%
\psdot(0,-1)%
\psdot(0.5,0.866)%
\psdot(-0.25,-0.067)%
\psline(-1.5,0)(-1,0)%
\psline(1.5,0)(1,0)%
\psline(0,-1)(-0.5,0.866)%
\psline(0,-1)(0.5,0.866)%
\rput[r](-0.9,0.7){\small 1}%
\rput[r](-0.9,-0.7){\small 2}%
\rput[r](-0.25,-0.5){\small 3}% 
\rput[r](-0.45,0.3){\small 3}% 
\rput[t](0,0.85){\small 4}% 
\rput[l](0.35,0){\small 6}% 
\rput[l](0.9,0.7){\small 7}% 
\rput[l](0.9,-0.7){\small 8}%
\end{center}
&
\begin{center}
\pscircle(0,0){1}%
\psdot(1,0)%
\psdot(-1,0)%
\psdot(-0.5,0.866)%
\psdot(0,-1)%
\psdot(0.5,0.866)%
\psdot(-0.707,-0.707)%
\psline(-1.5,0)(-1,0)%
\psline(1.5,0)(1,0)%
\psline(0,-1)(-0.5,0.866)%
\psline(0,-1)(0.5,0.866)%
\rput[r](-0.9,0.7){\small 1}%
\rput[r](-0.95,-0.45){\small 2}%
\rput[t](-0.5,-0.9){\small 2}%
\rput[r](-0.35,0){\small 3}% 
\rput[t](0,0.85){\small 4}% 
\rput[l](0.35,0){\small 6}% 
\rput[l](0.9,0.7){\small 7}% 
\rput[l](0.9,-0.7){\small 8}%
\end{center}
\\[0.6cm]
\centering $U_{\rm 6n2}$ & 
\centering $U_{\rm 7a1}$ & 
\centering $U_{\rm 7a2}$
\end{tabular}
\end{center}
\vspace{-2.5ex}
\mycaption{Master integral topologies with doubled propagators considered in this paper. 
The dot indicates a propagator that is raised to the power 2.
\label{fig:diag2}}
\end{figure}
%%%%%%%%%%%%%%%%%%%%%%%%%%%%%%%%%%%%%%%%%%%%%%%%%%%%%%%%%%%%%%

%%%%%%%%%%%%%%%%%%%%%%%%%%%%%%%%%%%%%%%%%%%%%%%%%%%%%%%%%%%%%%
\begin{figure}[tb]
\rule{0mm}{0mm}\\[-.2cm]
\begin{center}
\psset{linewidth=1pt}
\psset{dotsize=5pt}
\begin{tabular}{p{3.5cm}p{3.5cm}p{3.5cm}}
\begin{center}
\pscircle(0,0){1}%
\psdot(-1,0)%
\psdot(1,0)%
\psline(-1,0)(1,0)%
\rput[t](0,0.9){\small 1}% 
\rput[b](0,0.1){\small 2}%
\rput[b](0,-0.9){\small 3}% 
\end{center}
&
\begin{center}
\pscircle(0,0){1}%
\psdot(-1,0)%
\psdot(1,0)%
\psline(-1.5,0)(-1,0)%
\psline(1,0)(1.5,0)%
\psline(-1,0)(1,0)%
\rput[t](0,0.9){\small 2}% 
\rput[b](0,0.1){\small 3}%
\rput[b](0,-0.9){\small 4}% 
\end{center}
&
\begin{center}
\pscircle(0,0){1}%
\psdot(-1,0)%
\psdot(1,0)%
\psdot(0,1)%
\psline(-1.5,0)(-1,0)%
\psline(1,0)(1.5,0)%
\psline(-1,0)(1,0)%
\rput[r](-0.9,0.7){\small 2}%
\rput[l](0.9,0.7){\small 2}%
\rput[b](0,0.1){\small 3}%
\rput[b](0,-0.9){\small 4}% 
\end{center}
\\[0.6cm]
\centering $T_{\rm 3}$ & 
\centering $T_{\rm 3a}$ & 
\centering $T_{\rm 3a1}$  
\cr\\[0.3cm] 
\begin{center}
\pscircle(0,0){1}%
\psdot(1,0)%
\psdot(-1,0)%
\psdot(0,-1)%
\psline(-1.5,0)(-1,0)%
\psline(1,0)(1.5,0)%
\psarc(1,-1){1}{90}{180}%
\rput[r](-0.9,-0.6){\small 1}%
\rput[l](0.9,-0.6){\small 4}%
\rput[r](0.2,-0.2){\small 3}% 
\rput[t](0,0.9){\small 2}% 
\end{center}
&
\begin{center}
\pscircle(0,0){1}%
\psdot(1,0)%
\psdot(-1,0)%
\psdot(-0.707,-0.707)%
\psdot(0.707,-0.707)%
\psline(-1.5,0)(-1,0)%
\psline(1,0)(1.5,0)%
\psarc(0,-1.4){1}{45}{135}%
\rput[r](-1,-0.5){\small 1}%
\rput[l](1,-0.5){\small 1}%
\rput[b](0,-0.9){\small 4}%
\rput[b](0,-0.3){\small 3}% 
\rput[t](0,0.9){\small 2}% 
\end{center}
&
\begin{center}
\pscircle(0,0){1}%
\psdot(-1,0)%
\psdot(0,-1)%
\psdot(1,0)%
\psdot(0,1)%
\psline(-1.5,0)(-1,0)%
\psline(1,0)(1.5,0)%
\psline(0,1)(0,-1)%
\rput[r](-1,0.7){\small 1}% 
\rput[r](-1,-0.7){\small 2}%
\rput[l](1,0.7){\small 4}% 
\rput[l](1,-0.7){\small 5}%
\rput[l](0.15,0){\small 3}%
\end{center}
\\[0.6cm]
\centering $T_{\rm 4}$ & 
\centering $T_{\rm 4a1}$ & 
\centering $T_{\rm 5a}$
\end{tabular}
\end{center}
\vspace{-2.5ex}
\mycaption{Two-loop master integrals. 
\label{fig:2lmaster}}
\end{figure}
%%%%%%%%%%%%%%%%%%%%%%%%%%%%%%%%%%%%%%%%%%%%%%%%%%%%%%%%%%%%%%

%%%%%%%%%%%%%%%%%%%%%%%%%%%%%%%%%%%%%%%%%%%%%%%%%%%%%%%%%%%%%%

\section{Planar three-loop self-energy topologies and master integrals}
\label{sc:def}

\noindent
This article focuses on the evaluation of the ``planar-type'' three-loop self-energy
integrals that descend from diagrams containing two closed fermion loops. With ``planar-type'' we refer to topologies that can be considered
descendants of the ``master topology'' $U_{\rm 8a}$ in Fig.~\ref{fig:diag1},
by removing and/or doubling some propagators.

For the set of master integrals, we choose only integrals without numerator
terms, which generically are of the form
\begin{align}
U_{ij} &= i\frac{e^{3\gamma_{\rm E}\epsilon}}{\pi^{3D/2}}
  \int d^Dq_1\, d^Dq_2\, d^Dq_3 \;
  \frac{1}{[q_1^2-m_1^2]^{\nu_1} [(q_1+p)^2-m_2^2]^{\nu_2} 
           [(q_1-q_2)^2-m_3^2]^{\nu_3}} \notag \\
&\qquad\times \frac{1}{[q_2^2-m_4^2]^{\nu_4} [(q_2+p)^2-m_5^2]^{\nu_5}
                   [(q_2-q_3)^2-m_6^2]^{\nu_6} [q_3^2-m_7^2]^{\nu_7}
		   [(q_3+p)^2-m_8^2]^{\nu_8}} \label{eq:master}
\end{align}
Here $\epsilon = (4-D)/2$ and $D$ is the number of space-time dimensions in
dimensional regularization. Furthermore, the $\nu_k$ are integer numbers which
can be 0, 1 or 2 in our case.

To define our set of master integrals, we generated the diagrams with the topology
of $U_{\rm 8a}$ that occur in the three-loop self-energy diagrams with two
closed fermion loops using {\sc FeynArts 3} \cite{feynarts}. %\footnote{Specifically, we
%considered self-energy diagrams with two closed fermion loops in the Standard
%Model for this purpose.} 
We then performed an integral
reduction based on integration-by-parts identities with the help of FIRE 5
\cite{fire}. The resulting set of irreducible three-loop master integrals is
shown in Figs.~\ref{fig:diag1} and \ref{fig:diag2}. They are
all of the form given in eq.~\eqref{eq:master} with $\nu_k \in \{0,1,2\}$. 
There are additional
master integrals that factorize into products of one-loop and
two-loop integrals. The complete set of the latter is shown
in Fig.~\ref{fig:2lmaster} (see also Ref.~\cite{Weiglein:1993hd}).

We do not claim that this set of master integrals is minimal or optimal, but it
is suitable for numerical evaluation in terms of two-dimensional numerical
integrals, as will be demonstrated below. It should be emphasized that
additional master integrals are needed for diagrams that do not conform to the
planar master topology $U_{\rm 8a}$, such as
non-planar three-loop self-energy diagrams.

\medskip\noindent
The integrals in Figs.~\ref{fig:diag1} and \ref{fig:diag2} can be divided into
two groups: 
\begin{itemize}
\item Integrals with one- or two-loop sub-loop
self-energy. These can be evaluated efficiently using a dispersion relation for
the sub-loop bubbles \cite{disp2,disp2a,3lvac}.
\item Integrals without sub-loop self-energies. For these we employ a variant of
the method proposed in Ref.~\cite{Ghinculov:1996vd}.
This category comprises the
master integrals $U_{\rm 7a}$, $U_{\rm 8a}$, $U_{\rm 7a1}$ and $U_{\rm 7a2}$.
All the remaining master integrals belong to the former category.
\end{itemize}

%%%%%%%%%%%%%%%%%%%%%%%%%%%%%%%%%%%%%%%%%%%%%%%%%%%%%%%%%%%%%%

\section{Examples}
\label{sc:ex}

In the following subsections, our approaches for the numerical evaluation of the
master integrals are described in more detail for a few characteristic examples
from both categories.

\subsection{Double-bubble integrals: \boldmath $U_{\rm 5b}$}
\label{sc:u5b}

\noindent
A basic one-loop self-energy sub-loop can be expressed in terms of a dispersion
relation \cite{disp2}, $e.\,g.$
\begin{align}
\frac{e^{\gamma_{\rm E}\epsilon}}{i\pi^{D/2}} \int d^D q_1 \;
 \frac{1}{[q_1^2-m_a^2] [(q_1-q_2)^2-m_b^2]}
 \equiv B_0(q_2^2,m_a^2,m_b^2)
 = \int_0^\infty ds \; \frac{\Delta B_0(s,m_a^2,m_b^2)}{s-q_2^2-i\varepsilon}\,,
\end{align}
where $\Delta B_0$ is the discontinuity of the one-loop function $B_0$. In 
$D=4$ dimensions, it is given by
\begin{align}
\Delta B_0(s,m_a^2,m_b^2) &= \frac{1}{s}\sqrt{\lambda(s,m_a^2,m_b^2)} \,
\Theta\bigl(s-(m_a+m_b)^2\bigr)\,,
\end{align}
where $\lambda(x,y,z) = x^2+y^2+z^2-2(xy+yz+xz)$ and $\Theta(t)$ is the
Heaviside step function.

$U_{\rm 5b}$ contains two such sub-loop bubbles. Inserting the dispersion for
each of them, one obtains
\begin{align}
U_{\rm 5b} &= \int_0^\infty ds_1 \int_0^\infty ds_2 \;
 \Delta B_0(s_1, m_6^2, m_7^2) \, \Delta B_0(s_2, m_1^2, m_3^2) \notag \\
 &\qquad\times \frac{e^{\gamma_{\rm E}\epsilon}}{i\pi^{D/2}} \int d^D q_2 \;
 \frac{1}{[q_2^2-m_4^2][q_2^2-s_1][(q_2+p)^2-s_2]} \notag \displaybreak[0]
 \\[1ex]
&= \int_0^\infty ds_1 \int_0^\infty ds_2 \; 
 \Delta B_0(s_1, m_6^2, m_7^2) \, \Delta B_0(s_2, m_1^2, m_3^2)  \,
 \frac{B_0(p^2,s_1,s_2)-B_0(p^2,m_4^2,s_2^2)}{s_1-m_4^2-i\varepsilon} 
 \notag \displaybreak[0] \\[1ex]
&= \int_0^\infty ds_1 \int_0^\infty ds_2 \; 
 \Delta B_0(s_1, m_6^2, m_7^2) \, \Delta B_0(s_2, m_1^2, m_3^2)  \,
 \frac{B_0(p^2,s_1,s_2)}{s_1-m_4^2-i\varepsilon} \notag \\
 &\quad + B_0(m_4^2,m_6^2,m_7^2)\,T_{\rm 3a}(p^2,m_1^2,m_3^2)\,.
\label{eq:u5b_A}
\end{align}
Here $T_{\rm 3a}$ is a two-loop self-energy function, see
Fig.~\ref{fig:2lmaster}.

The $s_1$ and $s_2$ integrals in eq.~\eqref{eq:u5b_A} diverge at the upper
integral limit $\infty$, which can be attributed to the fact that $U_{\rm
5b}$ is UV divergent. The expression can be rendered UV finite by subtracting
suitable terms in the integrand that have the same UV singularity structure, but that
are otherwise simpler than the full $U_{\rm 5b}$ function. One way to achieve
this purpose is by subtracting the first two terms in a Taylor expansion in
$p^2$:
\begin{align}
&U_{\rm 5b}(p^2,m_2^2,m_3^2,m_4^2,m_6^2,m_7^2) = \notag \\
&\qquad U_{\rm 5b}(0,m_2^2,m_3^2,m_4^2,m_6^2,m_7^2)
 + p^2 U'_{\rm 5b}(0,m_2^2,m_3^2,m_4^2,m_6^2,m_7^2) \notag \\
&\qquad +B_0(m_4^2,m_6^2,m_7^2) \bigl[ T_{\rm 3a}(p^2,m_1^2,m_3^2)
 - T_{\rm 3a}(0,m_1^2,m_3^2) - p^2 T_{\rm 3a}(0,m_1^2,m_3^2) \bigr] \notag \\
&\qquad + \int_0^\infty ds_1 \int_0^\infty ds_2 \; 
 \Delta B_0(s_1, m_6^2, m_7^2) \, \Delta B_0(s_2, m_1^2, m_3^2) \notag \\
&\qquad\hspace{8em} \times
\frac{B_0(p^2,s_1,s_2)-B_0(0,s_1,s_2)-p^2B'_0(0,s_1,s_2)}{s_1-m_4^2-i\varepsilon}\,.
\label{eq:u5b_B}
\end{align}
Here the prime in $B'_0$ {\it etc.} denotes a derivative with respect to $p^2$.
The integral in the last two lines of eq.~\ref{eq:u5b_B} is now finite and can
be evaluated numerically. $U_{\rm 5b}(0,...)$ and $U'_{\rm 5b}(0,...)$ are
three-loop vacuum integrals, for which general methods for numerical evaluation
are known \cite{3lvac,3vil}. Similarly, the two-loop function $T_{\rm 3a}$
can be easily determined using the technique of Ref.~\cite{disp2}. The basic
one-loop function $B_0$ is known analytically \cite{c0} (see
Ref.~\cite{3lvac} for expressions that use the same conventions as in this
paper).

\subsection{Planar master topology: \boldmath $U_{\rm 8a}$}

\noindent
An simple method for numerically evaluating the master topology $U_{\rm 8a}$ was
presented in Ref.~\cite{Ghinculov:1996vd}.
This integral is UV finite and thus can be computed in four dimensions. It can
be written as
\begin{align}
U_{\rm 8a} = \int \frac{d^4 q}{i\pi^2}\; 
 \frac{C_0(p^2,(q+p)^2,q^2,m_1^2,m_2^2,m_3^2)
       C_0(q^2,p^2,(p+q)^2,m_6^2,m_7^2,m_8^2)}{[q^2 - m_4^2 + i\varepsilon]
       [(q+p)^2 - m_5^2 + i\varepsilon]}\,,
\end{align}
where $C_0$ is the basic one-loop vertex function, which is known analytically
in terms of logarithms and dilogarithms \cite{c0}. By moving to the
center-of-mass frame for $p$ and integrating over the solid angle of $\vec{q}$,
this becomes
\begin{align}
U_{\rm 8a} &= \frac{4\pi}{i\pi^2} \int_{-\infty}^\infty dq_0 \int_0^\infty
d|\vec{q}|\,|\vec{q}|^2\;
 \frac{C_0(p^2,y,x,m_1^2,m_2^2,m_3^2)
       C_0(x,p^2,y,m_6^2,m_7^2,m_8^2)}{[x - m_4^2 + i\varepsilon]
       [y - m_5^2 + i\varepsilon]}\,, 
\intertext{where}
x &= q^2 = q_0^2 - |\vec{q}|^2, \qquad
y = (q+p)^2 = q_0^2 - |\vec{q}|^2 + p^2 + 2q_0\sqrt{p^2}.
\end{align}
This formula suggests that it is convenient to adopt $x$ and $y$ as integration
variables, leading to the two-dimensional integral
\begin{align}
U_{\rm 8a} = \frac{1}{2i\pi p^2} \int_{-\infty}^\infty dx  \int_{-\infty}^\infty dy
\; &\sqrt{\lambda(x,y,p^2)}\,\Theta(\lambda(x,y,p^2)) \notag \\
&\times \frac{C_0(p^2,y,x,m_1^2,m_2^2,m_3^2)
       C_0(x,p^2,y,m_6^2,m_7^2,m_8^2)}{[x - m_4^2 + i\varepsilon]
       [y - m_5^2 + i\varepsilon]}\,. \label{eq:u8a}
\end{align}
Here the Heaviside $\Theta$ function is inserted to ensure that the integral
runs only over kinematically allowed values of $x$ and $y$.

The integrand in eq.~\eqref{eq:u8a} has singularities at $x=m_4^2$ and
$y=m_5^2$, which lead to difficulties for numerical integration routines. In
Ref.~\cite{Ghinculov:1996vd} this was addressed by using a deformation of the
integration contours into the complex plane. Here we instead split the
integrals into a residuum contribution and principal
value integral, according to the prescription \cite{3lvac}
\begin{align}
\int_{-\infty}^\infty dx \, \frac{f(x)}{x-\xi \pm i \varepsilon} =
\mp i\pi f(\xi) + \int_0^\infty dx' \, \frac{f(\xi+x')-f(\xi-x')}{x'}\,.
\label{eq:res}
\end{align}
This has the advantage that one does not need to worry about the complex contour
crossing any other singular points.

\subsection{Planar 7-propagator topology: \boldmath $U_{\rm 7a}$}

\noindent
In princple, the 7-propagator integral
\begin{align}
U_{\rm 7a} = \int \frac{d^4 q}{i\pi^2}\; \frac{C_0(p^2,(q+p)^2,q^2,m_1^2,m_2^2,m_3^2)C_0((p+q)^2,p^2,q^2,m_6^2,m_7^2,m_8^2)}{[q_1^2 - m_4^2 + i\varepsilon]}
\end{align}
can be treated with the same
approach as described for $U_{\rm 8a}$ in the previous sub-section.

However, it turns out that the $y$ integration is badly converging in this case.
A better convergence behavior is achieved by using $x$ and $q_0$ as integration
variables,
\begin{align}
U_{\rm 7a} = \frac{2}{i\pi} \int_{-\infty}^\infty dq_0 \int_{-\infty}^{q_0^2} dx \;
\frac{\sqrt{q_0^2-x}}{[x - m_4^2 + i\varepsilon]} \, &C_0(p^2,x+p^2+2q_0\sqrt{p^2},x,m_1^2,m_2^2,m_3^2)
 \notag \\
&\times C_0(x,p^2,x+p^2+2q_0\sqrt{p^2},m_6^2,m_7^2,m_8^2)\,. \label{eq:u7a}
\end{align}
For the $x$ integration again one can use the split into a residuum contribution and principal
value integral according to eq.~\eqref{eq:res}.

%%%%%%%%%%%%%%%%%%%%%%%%%%%%%%%%%%%%%%%%%%%%%%%%%%%%%%%%%%%%%%

\section{Implementation in TVID 2}
\label{sc:tvid}

\noindent
The TVID 2 package has two components:
\begin{itemize}
\item
One component runs in {\sc Mathematica} and performs the separation of the
master integrals into UV-divergent subtraction terms and finite remainder
functions, as described in section~\ref{sc:u5b} and appendix~\ref{sc:div}. This
separation can be performed algebraically (keeping the momenta and masses as
non-numerical symbols) or with numbers for the momenta and mass inserted from
the beginning (which will speed up the evaluation in {\sc Mathematica}).
\item
The second component carries out the numerical integration
of the finite remainder functions ($i.\,e.$ the functions labeled $U_{...,\rm
sub}$ in appendix~\ref{sc:div}). It is written in C and uses an adaptive
Gauss-Kronrod algorithm for the integrals, which yields a relative precision of
9--10 digits for most cases (see below for exceptions to this statement).
The input and output are handled through
simple text files that contain a list of numerical parameter values.
The {\sc Mathematica} component of TVID 2 can
directly call the numerical C component through an external system call.
\end{itemize}
The numerical of TVID uses quadruple precision floating points numbers to reduce
rounding-off errors in the tails of the integrals. However, for some master
integrals this
turns out not to be sufficient. For these cases, we use an asymptotic formula
for the integrand in the limit of large values of the dispersion variable $s_1$
and/or $s_2$. The asymptotic formula is used for values of $s_1+s_2 > s_{\rm
cut}$, with a suitably chosen value for the parameter $s_{\rm cut}$.
Specifically, $s_{\rm cut} = c\times p^2$, where $c$ is a constant that depends
on which function $U_{...,\rm sub}$ is being considered. Since $s_{\rm cut}$ is
proportional to $p^2$, there could be a loss of precision for cases when $p^2$
is either much larger or much smaller than the masses in the integral.

\medskip\noindent
The reader should take note of the following limitations of version 2.0 of 
TVID:
\begin{itemize}
\item The program cannot handle IR-divergent integrals. IR divergencies may
occur from certain configurations with multiple massless propagators or
threshold singularities. TVID 2.0 furthermore does not check whether a certain
parameter choice leads to an IR divergency; the user has to ensure that this is
the case.
\item There are additional cases ($i.\,e.$ particular combinations of input
parameters) that are IR finite but may require a special treatment within TVID
to avoid numerical instabilities. A few of these are implemented in version 2.0,
but there are probably many more that are currently missing. The authors encourage
users to submit any such special cases  when they discover them, and they will
be considered for implementation in future versions of TVID.
\item The finite remainder functions of $U_{\rm 6m1}$, $U_{\rm 6m3}$, $U_{\rm
6n2}$, $U_{\rm 7a1}$ and $U_{\rm 7a2}$ are related to those of $U_{\rm 6m}$,
$U_{\rm 6n}$ and $U_{\rm 7a}$ through mass
derivatives:
\begin{align}
U_{\rm 6m1,sub}(...) &= \tfrac{\partial}{\partial m_3^2}U_{\rm 6m,sub}(...),
&
U_{\rm 6m3,sub}(...) &= \tfrac{\partial}{\partial m_2^2}U_{\rm 6m,sub}(...), 
\displaybreak[0] \notag \\
U_{\rm 6n2,sub}(...) &= \tfrac{\partial}{\partial m_2^2}U_{\rm 6n,sub}(...),
\displaybreak[0] \notag \\
U_{\rm 7a1,sub}(...) &= \tfrac{\partial}{\partial m_3^2}U_{\rm 7a,sub}(...),
&
U_{\rm 7a2,sub}(...) &= \tfrac{\partial}{\partial m_2^2}U_{\rm 7a,sub}(...).
\end{align}
In version 2.0 of TVID these have been implemented using numerical differentiation
(based on a five-point stencil)\footnote{The reason for this choice is that the
reduction formula for the mass derivative $T_{5a}$ has rational coefficients
with high polynomial degrees in both the numerators and denominators, which
leads to many instabilities of the 0/0 type. Furthermore, the direct integration
of $U_{\rm 7a1}$ and $U_{\rm 7a2}$ involves very large numerical cancellations
between regions with positive and negative integrand, which leads to numerical
instabilities.}. As a result the delivered
precision for these functions is reduced to 6--7 digits.
\item Due to the presence of $C_0$ functions in the integrands of $U_{\rm 7a}$,
$U_{\rm 8a}$, $U_{\rm 7a1}$ and $U_{\rm 7a2}$, their evaluation is much more
time-consuming than that other master integrals. To mitigate this issue, TVID
2.0 uses only double precision floating point numbers for the evaluation of the
$C_0$ functions, and the target precision of these master integrals is reduced
to 8 digits for $U_{\rm 7a}$ and 6 digits for $U_{\rm 8a}$, $U_{\rm 7a1}$ and
$U_{\rm 7a2}$.
\item TVID 2.0 does not numerically evaluate the ${\cal O}(\epsilon)$ parts of
the two-loop functions in Fig.~\ref{fig:2lmaster} (labeled $T_{\rm ...,delta}$
in appendix~\ref{sc:div}). In principle, these ${\cal O}(\epsilon)$ terms are
needed if one wishes to evaluate all the master integrals in
Figs.~\ref{fig:diag1} and \ref{fig:diag2} to ${\cal O}(\epsilon^0)$. However, in
the calculation of any physical observable the $T_{\rm ...,delta}$ functions
should drop out when including the appropriate counterterm contributions, so
that their explicit numerical value should not be needed.
\item The algebraic part in TVID 2.0, which runs in {\sc Mathematica}, performs 
the separation of UV divergent subtraction terms
(as detailed in appendix~\ref{sc:div}) for individual master integrals or for
expressions that contain any linear combinations of these. However, the
resulting expressions can grow rather large, in particular if the masses are
treated symbolically at this stage. The {\sc Mathematica} code in TVID 2.0 
is not optimized to deal with very large expressions, and the user may have to
modify the {\tt PrepInt} method in TVID to avoid excessively long computing
times and related problems.
\end{itemize}
As a check and to calibrate the performance of TVID 2, we have performed 
comparisons with FIESTA 4.1 \cite{fiesta4}. If one takes each master integral in
isolation, the subtraction terms defined in appendix~\ref{sc:div} would require
the evaluation of some two-loop functions up to ${\cal O}(\epsilon)$. As already
mentioned above, these $T_{\rm ...,delta}$ functions are currently not
implemented in TVID 2.0, and they would also not be needed in the calculation of
physical quantities. To circumvent this issue, we have such defined modified
version of some master integrals, where certain terms have been subtracted that
reflect the physical counterterm structure. These modified master functions are
listed in Tab.~\ref{tab:idef}.
Note that the achievable numerical precision in some cases is somewhat reduced
due to cancellations between different terms in these
expressions.

%%%%%%%%%%%%%%%%%%%%%%%%%%%%%%%%%%%%%%%%%%%%%%%%%%%%%%%%%%%%%%
\begin{table}[!t]
\renewcommand{\arraystretch}{1.3}
\centering
\begin{tabular}{|l|}
\hline
$\widetilde{U}_{\rm 5a} = 
U_{\rm 5a}(p^2,1,3,5,6,7)-B_0(0,1,3)\,T_{\rm 3a}(p^2,5,6,7) 
 - B_0(0,6,7)\,T_{\rm 3a}(p^2,1,3,5)$\\
$\widetilde{U}_{\rm 5b} = 
U_{\rm 5b}(p^2,2,3,4,6,7)-B_0(0,6,7)\,T_{\rm 3a}(p^2,2,3,4)$\\
$\widetilde{U}_{\rm 5c} = 
U_{\rm 5c}(p^2,1,2,3,6,7)-[B_0(p^2,1,2)-B_0(0,1,2)]\,T_{\rm 3a}(1,3,6,7)$\\
$\widetilde{U}_{\rm 6a} = 
U_{\rm 6a}(p^2,1,3,4,5,6,7)-B_0(0,1,3)\,T_{\rm 4a}(p^2,4,5,6,7) 
 - B_0(0,6,7)\,T_{\rm 4a}(p^2,4,5,1,3)$ \\[-.5ex]
$\qquad\qquad - B_0(0,1,3)\,B_0(0,6,7)\,B_0(p^2,4,5)$\\
$\widetilde{U}_{\rm 6b} = 
U_{\rm 6b}(p^2,1,3,4,5,6,8)-B_0(0,1,3)\,T_{\rm 4a}(p^2,5,4,6,8) 
 - B_0(0,6,8)\,T_{\rm 4a}(p^2,4,5,1,3)$ \\[-.5ex]
$\qquad\qquad - B_0(0,1,3)\,B_0(0,6,8)\,B_0(p^2,4,5)$\\
$\widetilde{U}_{\rm 6c} = 
U_{\rm 6c}(p^2,1,2,3,4,6,7)-B_0(0,6,7)\,T_{\rm 4a}(p^2,1,2,3,4) 
 - B_0(p^2,1,2)\,T_{\rm 4a}(0,3,4,6,7)$ \\[-.5ex]
$\qquad\qquad - B_0(p^2,1,2)\,B_0(0,3,4)\,B_0(0,6,7)$\\
$\widetilde{U}_{\rm 6m} = 
U_{\rm 6m}(p^2,1,2,3,4,6,8)-B_0(0,6,8)\,T_{\rm 4a}(p^2,1,2,3,4) 
 - B_0(p^2,1,2)\,T_{\rm 4a}(0,3,4,6,8)$ \\[-.5ex]
$\qquad\qquad - B_0(p^2,1,2)\,B_0(0,3,4)\,B_0(0,6,8)$\\
$\widetilde{U}_{\rm 6n} = 
U_{\rm 6n}(p^2,1,2,3,6,7,8)
 - B_0(p^2,1,2)\,[T_{\rm 4a}(0,7,8,3,6) - B_0(0,7,8)\,B_0(0,3,6)]$ \\[-.5ex]
$\qquad\qquad - B_0(p^2,7,8)\,[T_{\rm 4a}(0,1,2,3,6) - B_0(0,1,2)\,B_0(0,3,6)]$ \\[-.5ex]
$\qquad\qquad - B_0(p^2,1,2)\,B_0(p^2,7,8)\,B_0(0,3,6)$\\
$\widetilde{U}_{\rm 5a1} = 
U_{\rm 5a1}(p^2,1,3,5,6,7)-B_0(0,1,3)\,T_{\rm 3a1}(p^2,5,6,7) 
 - B_0(0,6,7)\,T_{\rm 3a1}(p^2,5,1,3)$\\
$\widetilde{U}_{\rm 5a2} = 
U_{\rm 5a2}(p^2,1,3,5,6,7) - B_0(0,6,7)\,T_{\rm 3a1}(p^2,1,3,5)$\\
$\widetilde{U}_{\rm 5b1} = 
U_{\rm 5b1}(p^2,2,3,4,6,7)-B_0(0,6,7)\,T_{\rm 3a1}(p^2,4,2,3)$\\
$\widetilde{U}_{\rm 5b2} = 
U_{\rm 5b2}(p^2,2,3,4,6,7)-B_0(0,6,7)\,T_{\rm 3a1}(p^2,2,3,4)$\\
$\widetilde{U}_{\rm 5c1} = 
U_{\rm 5c1}(p^2,1,2,3,6,7)-B_0(p^2,1,2)\,T_{\rm 3a1}(0,7,6,3)$\\
$\widetilde{U}_{\rm 6m1} = 
U_{\rm 6m1}(p^2,1,2,3,4,6,8)-B_0(0,6,8)\,T_{\rm 4a3}(p^2,1,2,3,4)$ \\
$\widetilde{U}_{\rm 6m3} = 
U_{\rm 6m3}(p^2,1,2,3,4,6,8)-B_0(0,6,8)\,T_{\rm 4a2}(p^2,1,2,3,4)$ 
 \\[-.5ex]
$\qquad\qquad - B_{0,m_1}(p^2,2,1)\,T_{\rm 4a}(0,3,4,6,8) - B_{0,m_1}(p^2,1,2)\,B_0(0,3,4)\,B_0(0,6,8)$\\
$\widetilde{U}_{\rm 6n2} = 
U_{\rm 6n2}(p^2,1,2,3,6,7,8)
 - B_{0,m_1}(p^2,2,1)\,[T_{\rm 4a}(0,7,8,3,6)$ \\[-.5ex]
$\qquad\qquad  + B_0(p^2,7,8)\,B_0(0,3,6) - B_0(0,7,8)\,B_0(0,3,6)]$ \\
\hline
\end{tabular}
\mycaption{Definition of linear combinations of the master integrals, in which
all higher-order terms in $\epsilon$  of the basic integral functions 
(labeled $T_{\rm ...,delta}$ in appendix~\ref{sc:div}) drop out. These are used
in Tabs.~\ref{tab:comp1} and \ref{tab:comp2} for benchmarks and comparisons.
For the mass parameters the shorthand notations $1\equiv m_1^2$, {\it etc.} have
been employed. See appendix~\ref{sc:div} for the definition of the functions in the
subtraction terms.
\label{tab:idef}}
\end{table}
%%%%%%%%%%%%%%%%%%%%%%%%%%%%%%%%%%%%%%%%%%%%%%%%%%%%%%%%%%%%%%

The benchmark tests have been performed on a single core of an Intel Xeon CPU
with 3.7~GHz. Two parameter choices have been considered:
\begin{enumerate}
\item[a)] One choice where $p^2 < m_i^2$, so that $p^2$ is below any
threshold, and all master integrals are real. For these cases FIESTA has been
run with the settings\newline
{\tt\small CurrentIntegratorSettings =
\{\{"epsrel","1.000000E-05"\},\{"maxeval","5000000"\}\};\newline
ComplexMode = False;}\newline
The results are shown in Tab.~\ref{tab:comp1}.
\item[b)] A second choice where $p^2 \gg m_i^2$, all master integrals develop an
imaginary part. For these cases FIESTA has been
run with the settings\newline
{\tt\small CurrentIntegratorSettings =
\{\{"epsrel","1.000000E-04"\},\{"maxeval","5000000"\}\};\newline
ComplexMode = True;}\newline
The results are shown in Tab.~\ref{tab:comp2}.
\end{enumerate}
%%%%%%%%%%%%%%%%%%%%%%%%%%%%%%%%%%%%%%%%%%%%%%%%%%%%%%%%%%%%%%
\begin{table}[tbp]
\renewcommand{\arraystretch}{1.1}
$p^2=1.0, \,
m_1^2=1.1,\,
m_2^2=1.2,\,
m_3^2=1.3,\,
m_4^2=1.4,\,
m_5^2=1.5,\,
m_6^2=1.6,\,
m_7^2=1.7,\,
m_8^2=1.8$\\[1ex]
\centering
\begin{tabular}{|l|r@{.}l|l|r@{.}l|c|}
\hline
 & \multicolumn{3}{c|}{TVID 2.0} & \multicolumn{3}{c|}{FIESTA 4.1} \\
\cline{2-7}
 & \multicolumn{2}{c|}{Result} & Time [s] & \multicolumn{2}{c|}{Result} & Time [s] \\
\hline\hline
$U_{\rm 4a}$ & 38&7964435845(4) & \phantom{100}6.6 & 38&80(1) & 283 \\
$\widetilde{U}_{\rm 5a}$ & 9&828362321(2) & \phantom{100}0.5 & 9&830(2) & 283 \\
$\widetilde{U}_{\rm 5b}$ & 38&34202364(1) & \phantom{100}6.1 & 38&342(2) & 325 \\
$\widetilde{U}_{\rm 5c}$ & $-$2&97969664(6) & \phantom{100}9.4 & $-$2&980(2) & 354 \\
$\widetilde{U}_{\rm 6a}$ & 1&196967810(2) & \phantom{100}0.5 & 1&1970(1) & 315 \\
$\widetilde{U}_{\rm 6b}$ & 1&214272730(7) & \phantom{100}8.0 & 1&2143(1) & 314 \\
$\widetilde{U}_{\rm 6c}$ & $-$9&4490640(1) & \phantom{100}7.5 & $-$9&4491(1) & 340 \\
$\widetilde{U}_{\rm 6m}$ & $-$9&64795183(6) & \phantom{1}160 & $-$9&6480(1) & 336 \\
$\widetilde{U}_{\rm 6n}$ & $-$10&703719678(7) & \phantom{1}118 & $-$10&7037(2) & 365 \\
$\widetilde{U}_{\rm 7m}$ & 0&56501718077(4) & \phantom{10}78 & 0&56502(2) & 320 \\
$U_{\rm 7a}$ & $-$1&34380486(1)	& \phantom{1}206 & $-$1&34381(1) & 275 \\
$U_{\rm 8a}$ & 0&1224166(1) & \phantom{1}232 & 0&122418(1) & 542 \\
$U_{\rm 4a1}$ & $-$1&4651121210(1) & \phantom{100}1.5 & 1&465(3) & 163 \\
$U_{\rm 4a2}$ & $-$4&0102924343(4) & \phantom{100}3.0 & $-$4&0103(3) &
 \phantom{0}80 \\
$U_{\rm 4a3}$ & $-$3&1152647692(8) & \phantom{100}9.4 & $-$3&1153(4) & 
 \phantom{0}93 \\
$\widetilde{U}_{\rm 5a1}$ & 5&0248990852(4) & \phantom{100}0.5 &
 5&0248(1) & 164 \\
$\widetilde{U}_{\rm 5a2}$ & $-$3&3851828312(5) & \phantom{100}0.5 &
 $-$3&3852(2) & 156 \\
$\widetilde{U}_{\rm 5b1}$ & 7&419421372(3) & \phantom{10}13.5 &
 7&4194(1) & 166 \\
$\widetilde{U}_{\rm 5b2}$ & $-$3&261463313(4) & \phantom{10}30 &
 $-$3&2615(2) & 153 \\
$\widetilde{U}_{\rm 5c1}$ & $-$4&0173586528(4) & \phantom{10}30 &
 $-$4&0174(2) & 165 \\
$\widetilde{U}_{\rm 6m1}$ & 0&74392431(2) & \phantom{1}638 &
 0&74392(3) & 148 \\
$U_{\rm 6m2}$ & $-$1&33361342263(2) & \phantom{10}12.7 & $-$1&33362(1) & 105 \\
$\widetilde{U}_{\rm 6m3}$ & $-$0&1300547(6) & \phantom{1}623 &
 $-$0&1301(1) & 195 \\
$U_{\rm 6n1}$ & $-$1&63165820287(4) & \phantom{100}7.5 & $-$1&63166(1) & 100 \\
$\widetilde{U}_{\rm 6n2}$ & 0&36932150(7) & \phantom{1}444 & 0&3693(1) & 175 \\
$U_{\rm 7a1}$ & 0&101053(1) & \phantom{1}812 & 0&101054(1) & 264 \\
$U_{\rm 7a2}$ & 0&220078(1) & \phantom{1}822 & 0&220080(3) & 269 \\
\hline
\end{tabular}
\mycaption{Comparison of results between TVID 2.0 and FIESTA 4.1 \cite{fiesta4}.
Where applicable, the linear combinations defined in Tab.~\ref{tab:idef} are
used, indicated by the tilde ($\widetilde{U}_{xxx}$).
The table lists the finite part of a series expansion in $\epsilon$, with the
numbers in brackets giving the integration error in the last quoted digit. Also
given are the run times for the numerical integration of the two programs, which
exclude the preparation time (in {\sc Mathematica}) of the integrals in either
case.
\label{tab:comp1}}
\end{table}
%%%%%%%%%%%%%%%%%%%%%%%%%%%%%%%%%%%%%%%%%%%%%%%%%%%%%%%%%%%%%%
\begin{table}[tb]
\renewcommand{\arraystretch}{1.1}
$p^2=40, \,
m_1^2=1.1,\,
m_2^2=1.2,\,
m_3^2=1.3,\,
m_4^2=1.4,\,
m_5^2=1.5,\,
m_6^2=1.6,\,
m_7^2=1.7,\,
m_8^2=1.8$\\[1ex]
\centering
\begin{tabular}{|l|r@{.}l|l|r@{.}l|c|}
\hline
 & \multicolumn{3}{c|}{TVID 2.0} & \multicolumn{3}{c|}{FIESTA 4.1} \\
\cline{2-7}
 & \multicolumn{2}{c|}{Result} & Time [s] & \multicolumn{2}{c|}{Result} & Time [s] \\
\hline\hline
$U_{\rm 4a}$ & $-$149&6944621(5) & \phantom{10}17.6 & 
	$-$149&7(1) & 3052 \\[-.5ex]
	& +9&6099138(5)$\,i$ && +9&6(1)$\,i$ & \\
$\widetilde{U}_{\rm 5a}$ & 53&705925142(1) & 
	\phantom{100}0.5 & 53&71(8) & 2865 \\[-.5ex]
	& $-$20&874552008(1)$\,i$ && $-$20&88(8)$\,i$ & \\
$\widetilde{U}_{\rm 5b}$ & 91&63152677(6)  & 
	\phantom{10}15.3 & 91&64(7)  & 2826 \\[-.5ex]
	& $-$2&54536001(6)$\,i$ && $-$2&54(7)$\,i$ & \\
$\widetilde{U}_{\rm 5c}$ & $-$18&763016(2)  & 
	\phantom{10}18.4 & $-$18&8(2) & 3728 \\[-.5ex]
	& +0&452121(2)$\,i$ && +0&5(2)$\,i$ & \\
$\widetilde{U}_{\rm 6a}$ & 3&347688278(2) & 
	\phantom{100}0.6 & 3&35(2) & 5216 \\[-.5ex]
	& $-$2&796453548(2)$\,i$ && $-$2&79(2)$\,i$ & \\
$\widetilde{U}_{\rm 6b}$ & 3&461079863(7)  & 
	\phantom{10}16.1 & 3&46(1) & 5300 \\[-.5ex]
	& $-$0&420922147(7)$\,i$ && $-$0&42(1)$\,i$ & \\
$\widetilde{U}_{\rm 6c}$ & $-$8&7387474(1) & 
	\phantom{10}15.0 & $-$8&74(1) & 5549 \\[-.5ex]
	& $-$0&3410452(1)$\,i$ && $-$0&34(1)$\,i$ & \\
$\widetilde{U}_{\rm 6m}$ & $-$11&094545131(6) &
	\phantom{1}989 & $-$11&094(7) & 5585 \\[-.5ex]
	& +4&390391111(6)$\,i$ && +4&391(7)$\,i$ & \\
$U_{\rm 7a}$ & 0&572024801(5) & \phantom{1}174 &
	0&57186(5) & 6116 \\[-.5ex]
	& $-$0&361496849(5)$\,i$ && $-$0&36139(4)$\,i$ & \\
$U_{\rm 8a}$ & 0&01238717(2) & \phantom{1}253 & 0&012353(3) & 11407 \\[-.5ex]
	& $-$0&16344185(2)$\,i$ && $-$0&016361(3)$\,i$ & \\
\hline
\end{tabular}
\mycaption{Same as Tab.~\ref{tab:comp1}, but for a larger value of $p^2$.
\label{tab:comp2}}
\end{table}
%%%%%%%%%%%%%%%%%%%%%%%%%%%%%%%%%%%%%%%%%%%%%%%%%%%%%%%%%%%%%%
As can be seen from the tables, there is generally excellent agreement between
TVID 2 and FIESTA within integration errors. 
Only for the cases with 7 or more propagators ($U_{\rm 7a}$ and
$U_{\rm 8a}$) that require contour deformation in sector decomposition
(Tab.~\ref{tab:comp2}), the discrepancy between the two programs is larger than
the integration error reported by FIESTA. This may in part be due to the contour
deformation being unable to make the integrands sufficiently smooth in those
cases. 

TVID 2 generally achieves 6--10
digit precision within run times ranging from less than 1 second to about 20
minutes. The precision and run time are not crucially affected by the presence of
physical cuts ($i.\,e.$ whether $p^2$ is below or above any of the thresholds of
the integral). 
Note that the run times shown in the tables only reflect the time for the
numerical integrations. Additionally, the preparation time for the integrals in
the {\sc Mathematica} module of FIESTA can be significant, in particular for the
cases that require contour deformation (Tab.~\ref{tab:comp2})\footnote{In fact,
the preparation of the contour deformation for $U_{\rm 8a}$ in FIESTA takes
several days.}.

%%%%%%%%%%%%%%%%%%%%%%%%%%%%%%%%%%%%%%%%%%%%%%%%%%%%%%%%%%%%%%

\section{Conclusions}
\label{sc:concl}

\noindent
Numerical integration is currently the most efficient way to evaluate the finite
pieces of multi-loop integrals with arbitrary masses, which set a multitude of
different scales. The program TVID 2 aims to provide an efficient and
automizable procedure for the numerical evaluation of a large class of three-loop self-energy
integrals.

The master topologies fall into one of two categories. Topologies containing one
or two sub-loop self-energies are generally UV-divergent and are treated by
subtracting simpler, known integrals with the same divergence structure. The
remaining finite pieces are written as one- or two-dimensional integrals over
analytically known functions with the help of dispersion relations, which in
turn can be evaluated numerically. The topologies without sub-loops
self-energies are UV-finite and can be performed as two-dimensional integrals
containing one-loop triangle functions in integrand.  The general procedure for
this class has been described in Ref.~\cite{Ghinculov:1996vd} and was adapted to
avoid complex contour deformation. Several technical subtleties when following
this approach are discussed in this paper.
TVID~2 contains also contains all the basic elements of the S2LSE package for
two-loop self-energy master integrals \cite{baubergerdiss}.

In order to confirm the correctness and accuracy of the implementation we
carried out a multitude of independent comparisons. To that end we utilized the
publicly available package FIESTA and found excellent agreement for almost
all the master integrals. 
If the external momentum squared is sufficiently large
so that the master integrals develop a non-zero imaginary part, the precision
and accuracy of FIESTA is significantly diminished, leading to less perfect
agreement between the two programs in some cases. Generally, 
TVID 2 achieves 6-10 digit precision for all the master integrals 
in run times of seconds to minutes.

At the present time a few issues remain to be addressed before one has
complete computational control over the entirety of three-loop self-energy-type
integrals. The current version of TVID is not equipped to treat integrals that
exhibit soft/collinear divergencies, which can occur in some master integrals
for
certain input parameter combinations. Furthermore, even though we cover a large
subset of master integrals, there is a number of topologies missing, specifically
the descendants of the ``Mercedes star'' and the non-planar topologies.
These issues are delegated to future work.

%%%%%%%%%%%%%%%%%%%%%%%%%%%%%%%%%%%%%%%%%%%%%%%%%%%%%%%%%%%%%%

\section*{Acknowledgments}

\noindent
The authors acknowledge contributions by A.~Haidet at early stages of the
project.
The work of A.~F.\ has been supported in part by the National Science Foundation under
grant no.\ PHY-1820760.
D.~W.\ is supported by the DOE grants DE-FG02-91ER40684 and DE-AC02-06CH11357. Part of this research used resources of the Argonne Leadership Computing Facility, which is a DOE Office of Science User Facility supported under Contract DE-AC02-06CH11357.
A.~F.\ is grateful to J.~Gluza and other members of the Polish 
National Science Centre grant no.~2017/25/B/ST2/01987 for  
discussions and exchanges related to this topic.

%%%%%%%%%%%%%%%%%%%%%%%%%%%%%%%%%%%%%%%%%%%%%%%%%%%%%%%%%%%%%%

%\newpage

\appendix
\section{Subtraction of divergent terms}
\label{sc:div}

\noindent
Before the master integrals in Figs.~\ref{fig:diag1}, \ref{fig:diag2} and
\ref{fig:2lmaster} can be evaluated numerically, one needs to remove their UV
divergencies. This can be achieved by subtracting simpler integrals that have
the same UV singularity structure, but that are known in the literature. The
finite remainder parts are denoted by $T_{xxx,\rm sub}$ and $U_{xxx,\rm sub}$ in
the equations below. These can be evaluated with the numerical part of TVID 2.

For the sake of brevity, the following shorthand notations are used in this
section:
\begin{align}
B_0(p^2,1,2) &\equiv B_0(p^2,m_1^2,m_2^2), \text{ etc.} \\
B_{0,m_1}(p^2,1,2) &= \tfrac{\partial}{\partial m_1^2}B_0(p^2,1,2)
\intertext{$B_0^{(n)}$ denotes the $B_0$ function with the
order-$n$ Taylor expansion subtracted,}
B_0^{(n)}(p^2,1,2) &= B_0(p^2,1,2) - \sum_{k=0}^n \frac{p^{2k}}{k!}\,
\frac{\partial^k}{\partial(p^2)^k}B_0(p^2,1,2)\Big|_{p^2=0}.
\end{align}
$T_{\rm 3a}^{(n)}$, $T_{\rm 4a}^{(n)}$ and $T_{\rm 5a}^{(n)}$ are defined in a similar fashion.

The discontinuities of $B_0$ and $B_{0,m_1}$ are given by, in $D=4$ dimensions,
\begin{align}
\Delta B_0(s,m_a^2,m_b^2) &= \frac{1}{s}\sqrt{\lambda(s,m_a^2,m_b^2)} \,
\Theta\bigl(s-(m_a+m_b)^2\bigr)\,, \\
\Delta B_{0,m_1}(s,m_a^2,m_b^2) &= \frac{m_a^2-m_b^2-s}{s\,\sqrt{\lambda(s,m_a^2,m_b^2
)}} \,
\Theta\bigl(s-(m_a+m_b)^2\bigr)\,,
\intertext{where}
\lambda(x,y,z) &= x^2+y^2+z^2-2(xy+yz+xz)\,, \label{eq:lambda}
\end{align}
and $\Theta(t)$ is the Heaviside step function.

\medskip\noindent
The UV subtractions for the two-loop integrals can be taken over from
Ref.~\cite{baubergerdiss}:
\begin{align}
T_{\rm 3a}(p^2,2,3,4) &= T_3(2,3,4) + T'_{\rm 3a}(0,2,3,4) \notag \\
 &\quad + T_{\rm 3a,sub}(p^2,2,3,4) + \epsilon \, T_{\rm 3a,delta}(p^2,2,3,4) +
 {\cal O}(\epsilon^2)\,, \\
T_{\rm 3a,sub}(p^2,2,3,4) &= -\int_0^\infty ds \; \Delta B_0(s,2,3) \,
 B_0^{(1)}(p^2,s,4) 
 \,, \displaybreak[0]\\[1ex]
T_{\rm 3a1}(p^2,2,3,4) &= T_{3,m_1}(2,3,4)
 + T_{\rm 3a1,sub}(p^2,2,3,4)+ T_{\rm 3a1,delta}(p^2,2,3,4) +
 {\cal O}(\epsilon^2)\,, \\
T_{\rm 3a1,sub}(p^2,2,3,4) &= -\int_0^\infty ds \; \Delta B_{0,m_1}(s,2,3) \,
 B_0^{(0)}(p^2,s,4) 
 \,, \displaybreak[0]\\[1ex]
T_{\rm 4a}(p^2,1,2,3,4) &= T_{\rm 4a}(0,1,2,3,4) + B_0(1,3,4)\,B_0^{(0)}(p^2,1,2)
 \notag \\
 &\quad + T_{\rm 4a,sub}(p^2,1,2,3,4) + \epsilon \, T_{\rm 4a,delta}(p^2,1,2,3,4) +
 {\cal O}(\epsilon^2)\,, \\
T_{\rm 4a,sub}(p^2,1,2,3,4) &= -\int_0^\infty ds \; \frac{\Delta
 B_0(s,3,4)}{s-m_1^2-i\varepsilon}\, B_0^{(0)}(p^2,s,2) 
 \,, \displaybreak[0]\\[1ex]
T_{\rm 4a1}(p^2,1,2,3,4) &= B_0(1,3,4)\,B_{0,m_1}(p^2,1,2)
 \notag \\
 &\quad + T_{\rm 4a1,sub}(p^2,1,2,3,4) + \epsilon \, T_{\rm 4a1,delta}(p^2,1,2,3,4) +
 {\cal O}(\epsilon^2)\,, \\
T_{\rm 4a1,sub}(p^2,1,2,3,4) &= -\int_0^\infty ds \; \frac{\Delta
 B_0(s,3,4)}{(s-m_1^2-i\varepsilon)^2}\, [B_0(p^2,s,2)-B_0(p^2,1,2)] \,.
\end{align}
The two-loop vacuum integral $T_3$ is known analyically \cite{2lvac,2lvaca}, and
$T'_{\rm 3a}(0,...)$ and $T_{\rm 4a}(0,...)$ can be reduced to linear combinations
of $T_3$ functions and one-loop functions 
by using partial fractioning and integration-by-parts
relations\footnote{\label{note1}Explicit formulae are included in the {\sc
Mathematica} part of
TVID 2.}. Here $T'_{\rm 3a}$ denotes the derivative of $T_{\rm 3a}$ with
respect to $p^2$.

\medskip\noindent
In a similar fashion, the UV subtraction for the three-loop
self-energy integrals $U_{xxx}$ leads to
the functions $U_{xxx}(0,...)$, $U'_{\rm xxx}(0,...)$ and $U''_{\rm
xxx}(0,...)$,
which are three-loop vacuum integrals.
They also can be reduced to basic master vacuum integrals
\cite{3lvac,3vil} with the help of partial fractioning and integration by
parts\footnoteref{note1}. As before, $U'_{\rm xxx}$ and $U''_{\rm xxx}$ denote the first and second
derivative of $U_{xxx}$ with respect to $p^2$.
\begin{align}
U_{\rm 4a}(p^2,1,3,6,8) &= U_{\rm 4a}(0,1,3,6,8) 
 + p^2 U'_{\rm 4a}(0,1,3,6,8) + \frac{p^4}{2} U''_{\rm 4a}(0,1,3,6,8) \notag \\
&\quad +U_{\rm 4a,sub}(p^2,1,3,6,8) \,,\\[.5ex]
U_{\rm 4a,sub}(...) &=
 \int_0^\infty ds_1 \int_0^\infty ds_2 \; \Delta B_0(s_1,1,3)\,
  \Delta B_0(s_2,6,8) \, B_0^{(2)}(p^2,s_1,s_2) 
\,, \displaybreak[0]\\[1ex]
U_{\rm 5a}(p^2,1,3,5,6,7) &= U_{\rm 5a}(0,1,3,5,6,7) 
 + p^2 U'_{\rm 5a}(0,1,3,5,6,7) \notag \\
&\quad +B_0(0,1,3)\, T^{(1)}_{\rm 3a}(p^2,5,6,7)
 +B_0(0,6,7)\, T^{(1)}_{\rm 3a}(p^2,1,3,5) \notag \\
&\quad +U_{\rm 5a,sub}(p^2,1,3,5,6,7) \,,\\[.5ex]
U_{\rm 5a,sub}(...) &=
 -\int_0^\infty ds \; \Delta B_0(s,1,3)\,
  B_0^{(0)}(s,6,7) \, B_0^{(1)}(p^2,s,5) \notag \\
&\quad -\int_0^\infty ds \; \Delta B_0(s,6,7) \,
  B_0^{(0)}(s,1,3) \, B_0^{(1)}(p^2,s,5)
\,, \displaybreak[0]\\[1ex]
U_{\rm 5b}(p^2,2,3,4,6,7) &= U_{\rm 5b}(0,2,3,4,6,7) 
 + p^2 U'_{\rm 5b}(0,2,3,4,6,7)  \notag \\
&\quad +B_0(4,6,7)\, T^{(1)}_{\rm 3a}(p^2,2,3,4) \notag \\
&\quad +U_{\rm 5b,sub}(p^2,2,3,4,6,7) \,,\\[.5ex]
U_{\rm 5b,sub}(...) &=
 \int_0^\infty ds_1\int_0^\infty ds_2 \; 
 \frac{\Delta B_0(s_1,6,7)}{s_1-m_4^2-i\varepsilon}\;
 \Delta B_0(s_2,2,3)\, B_0^{(1)}(p^2,s_1,s_2) 
\,, \displaybreak[0]\\[1ex]
U_{\rm 5c}(p^2,1,2,3,6,7) &= U_{\rm 5c}(0,1,2,3,6,7) 
 + p^2 U'_{\rm 5c}(0,1,2,3,6,7) \notag \\
&\quad +T_{\rm 3a}(1,3,6,7) \, B_0^{(1)}(p^2,1,2) \notag \\
&\quad +U_{\rm 5c,sub}(p^2,1,2,3,6,7) \,,\\[.5ex]
U_{\rm 5c,sub}(...) &=
 \int_0^\infty ds_1\int_0^\infty ds_2 \; 
 \frac{\Delta B_0(s_1,s_2,3)}{s_1-m_1^2-i\varepsilon}\;
 \Delta B_0(s_2,6,7) \, B_0^{(1)}(p^2,s_1,2) 
\,, \displaybreak[0]\\[1ex]
U_{\rm 6a}(p^2,1,3,4,5,6,7) &= U_{\rm 6a}(0,1,3,4,5,6,7) \notag \\
&\quad +B_0(4,1,3)\, T^{(0)}_{\rm 4a}(p^2,4,5,6,7)  
 +B_0(4,6,7)\, T^{(0)}_{\rm 4a}(p^2,4,5,1,3) \notag \\
&\quad -B_0(4,1,3)\, B_0(4,6,7)\, B_0^{(0)}(p^2,4,5) \notag \\
&\quad +U_{\rm 6a,sub}(p^2,1,3,4,5,6,7) \,,\\[.5ex]
U_{\rm 6a,sub}(...) &=
 -\int_0^\infty ds \; \frac{\Delta B_0(s,1,3)}{s-m_4^2-i\varepsilon}\,
  [B_0(s,6,7)-B_0(4,6,7)]\, B_0^{(0)}(p^2,s,5) \notag \\
&\quad -\int_0^\infty ds \; \frac{\Delta B_0(s,6,7)}{s-m_4^2-i\varepsilon}\,
  [B_0(s,1,3)-B_0(4,1,3)]\, B_0^{(0)}(p^2,s,5)
\,, \displaybreak[0]\\[1ex]
U_{\rm 6b}(p^2,1,3,4,5,6,8) &= U_{\rm 6b}(0,1,3,4,5,6,8) \notag \\
&\quad +B_0(4,1,3)\, T^{(0)}_{\rm 4a}(p^2,5,4,6,8) 
 +B_0(5,6,8)\, T^{(0)}_{\rm 4a}(p^2,4,5,1,3) \notag \\
&\quad -B_0(4,1,3)\, B_0(5,6,8)\, B_0^{(0)}(p^2,4,5) \notag \\
&\quad +U_{\rm 6b,sub}(p^2,1,3,4,5,6,8) \,,\\[.5ex]
U_{\rm 6b,sub}(...) &=
 \int_0^\infty ds_1\int_0^\infty ds_2 \; 
 \frac{\Delta B_0(s_1,1,3)}{s_1-m_4^2-i\varepsilon}\;
 \frac{\Delta B_0(s_2,6,8)}{s_2-m_5^2-i\varepsilon}\,
  B_0^{(0)}(p^2,s_1,s_2) 
\,, \displaybreak[0]\\[1ex]
U_{\rm 6c}(p^2,1,2,3,4,6,7) &= U_{\rm 6c}(0,1,2,3,4,6,7) \notag \\
&\quad +[T_{\rm 4a}(1,4,3,6,7) - B_0(1,3,4)\, B_0(4,6,7) ]\, B_0^{(0)}(p^2,1,2) \notag \\
&\quad +B_0(4,6,7)\, T^{(0)}_{\rm 4a}(p^2,1,2,3,4) \notag \\
&\quad +U_{\rm 6c,sub}(p^2,1,2,3,4,6,7) \,,\\[.5ex]
U_{\rm 6c,sub}(...) &=
 \int_0^\infty ds_1\int_0^\infty ds_2 \; 
 \frac{\Delta B_0(s_1,s_2,3)}{s_1-m_1^2-i\varepsilon}\;
 \frac{\Delta B_0(s_2,6,7)}{s_2-m_4^2-i\varepsilon}\,
  B_0^{(0)}(p^2,s_1,2) 
\,, \displaybreak[0]\\[1ex]
U_{\rm 6m}(p^2,1,2,3,4,6,8) &= U_{\rm 6m}(0,1,2,3,4,6,8) \notag \\
&\quad +B_0(0,6,8) \, T^{(0)}_{\rm 4a}(p^2,1,2,3,4)
 + T_{\rm 4a}(0,0,0,6,8) \, B_0^{(0)}(p^2,1,2) \notag \\
&\quad + U_{\rm 6m,sub}(p^2,1,2,3,4,6,8)  \,,\\[.5ex]
U_{\rm 6m,sub}(...) &= -\int ds \; \Delta B_0(s,6,8)
 \Bigl [ T_{\rm 5a}^{(0)}(p^2,1,2,3,4,s) \notag \\
&\qquad\qquad - \frac{1}{s} \bigl [
  B_0(0,0,s) \, B_0^{(0)}(p^2,1,2) - T_{\rm 4a}^{(0)}(p^2,1,2,3,4) \bigr ] 
  \Bigr ]
\,, \displaybreak[0]\\[1ex]
U_{\rm 6n}(p^2,1,2,3,6,7,8) &= U_{\rm 6n}(0,1,2,3,6,7,8) \notag \\
&\quad +B_0(0,3,6) \, [B_0(p^2,1,2)\,B_0(p^2,7,8) - B_0(0,1,2)\,B_0(0,7,8)]
 \notag \\
&\quad + T_{\rm 4a}(0,0,0,3,6) \, [B_0^{(0)}(p^2,1,2) + B_0^{(0)}(p^2,7,8)] \notag \\
&\quad + U_{\rm 6n,sub}(p^2,1,2,3,6,7,8)  \,,\\[.5ex]
U_{\rm 6n,sub}(...) &= -\int ds \; \Delta B_0(s,3,6)
 \Bigl [ T_{\rm 5a}^{(0)}(p^2,1,2,s,7,8) \notag \\
&\qquad - \frac{1}{s} \bigl [
  B_0(0,0,s) \, \bigl ( B_0^{(0)}(p^2,1,2) + B_0^{(0)}(p^2,7,8) \bigr ) \notag \\ 
&\qquad\qquad  - B_0(p^2,1,2)\,B_0(p^2,7,8) + B_0(0,1,2)\,B_0(0,7,8) \bigr ] 
  \Bigr ]
\,, \displaybreak[0]\\[1ex]
U_{\rm 7m}(p^2,1,2,3,4,5,6,8) &= B_0(5,6,8) \, T_{\rm 5a}(p^2,1,2,3,4,5) +
 U_{\rm 7m,sub}(p^2,1,2,3,4,5,6,8)  \,,\\[.5ex]
U_{\rm 7m,sub}(...) &= -\int_0^\infty ds \; 
 \frac{\Delta B_0(s,6,8)}{s_1-m_5^2-i\varepsilon}\, T_{\rm 5a}(p^2,1,2,3,4,s)
\,. 
\end{align}
$U_{\rm 7a}$ and $U_{\rm 8a}$ are UV finite.
\begin{align}
U_{\rm 4a1}(p^2,1,3,6,8) &= \tfrac{\partial}{\partial m_1^2}U_{\rm 4a}(0,1,3,6,8) 
 + p^2 \,\tfrac{\partial}{\partial m_1^2}U'_{{\rm 4a},m_1}(0,1,3,6,8) \notag \\
&\quad +U_{\rm 4a1,sub}(p^2,1,3,6,8) \,,\\[.5ex]
U_{\rm 4a1,sub}(...) &=
 \int_0^\infty ds_1 \int_0^\infty ds_2 \; \Delta B_{0,m_1}(s_1,1,3)\,
  \Delta B_0(s_2,6,8) \, B_0^{(1)}(p^2,s_1,s_2) 
\,, \displaybreak[0]\\[1ex]
U_{\rm 4a2}(p^2,1,3,6,8) &= 
 \tfrac{\partial^2}{\partial m_1^2 \partial m_3^2}U_{\rm 4a}(0,1,3,6,8) 
 + U_{\rm 4a2,sub}(p^2,1,3,6,8) \,,\\[.5ex] 
U_{\rm 4a2,sub}(...) &=
 \int_0^\infty ds_1 \int_0^\infty ds_2 \; \Delta B_{0,m_1}(s_1,1,6)\,
  \Delta B_{0,m_1}(s_2,3,8) \, B_0^{(1)}(p^2,s_1,s_2) 
\,, \displaybreak[0]\\[1ex]
U_{\rm 4a3}(p^2,1,3,6,8) &= 
 \tfrac{1}{2}\,\tfrac{\partial^2}{\partial (m_1^2)^2}U_{\rm 4a}(0,1,3,6,8) 
 + \tfrac{p^2}{2}\,\tfrac{\partial^2}{\partial (m_1^2)^2} U'_{\rm 4a}(0,1,3,6,8) \notag \\
&\quad +U_{\rm 4a3,sub}(p^2,1,3,6,8) \,,\\[.5ex]
U_{\rm 4a3,sub}(...) &=
 \frac{1}{2}\int_0^\infty ds_1 \int_0^\infty ds_2 \; \Delta K_3(s_1,1,3)\,
  \Delta B_0(s_2,6,8) \, B_{0,m_1}^{(1)}(p^2,s_1,s_2)\,, \notag \\
&\text{where } K_3(s,m_a^2,m_b^2) = \frac{s-m_a^2-m_b^2}{m_a^2 \,
 \sqrt{\lambda(s,m_a^2,m_b^2)}} \, \Theta\bigl(s-(m_a+m_b)^2\bigr)
\,, \displaybreak[0]\\[1ex]
U_{\rm 5a1}(p^2,1,3,5,6,7) &= \tfrac{\partial}{\partial m_5^2}U_{\rm 5a}(0,1,3,5,6,7) 
 \notag \\
&\quad +B_0(0,1,3)\, T^{(0)}_{\rm 3a1}(p^2,5,6,7)
 +B_0(0,6,7)\, T^{(0)}_{\rm 3a1}(p^2,5,1,3) \notag \\
&\quad +U_{\rm 5a1,sub}(p^2,1,3,5,6,7) \,,\\[.5ex]
U_{\rm 5a1,sub}(...) &=
 -\int_0^\infty ds \; \Delta B_0(s,1,3)\,
  B_0^{(0)}(s,6,7) \, B_{0,m_1}^{(0)}(p^2,5,s) \notag \\
&\quad -\int_0^\infty ds \; \Delta B_0(s,6,7) \,
  B_0^{(0)}(s,1,3) \, B_{0,m_1}^{(0)}(p^2,5,s)
\,, \displaybreak[0]\\[1ex]
U_{\rm 5a2}(p^2,1,3,5,6,7) &= \tfrac{\partial}{\partial m_1^2}U_{\rm 5a}(0,1,3,5,6,7) 
  +B_0(0,6,7)\, T^{(0)}_{\rm 3a1}(p^2,1,3,5) \notag \\
&\quad +U_{\rm 5a2,sub}(p^2,1,3,5,6,7) \,,\\[.5ex]
U_{\rm 5a2,sub}(...) &=
 -\int_0^\infty ds \; \Delta B_{0,m_1}(s,1,3)\,
  B_0^{(0)}(s,6,7) \, B_0^{(0)}(p^2,5,s) \notag \\
&\quad -\int_0^\infty ds \; \Delta B_0(s,6,7) \,
  B_{0,m_1}^{(0)}(s,1,3) \, B_0^{(0)}(p^2,5,s)
\,, \displaybreak[0]\\[1ex]
U_{\rm 5b1}(p^2,2,3,4,6,7) &= \tfrac{\partial}{\partial m_4^2}U_{\rm 5b}(0,2,3,4,6,7) 
 +B_0(4,6,7)\, T^{(0)}_{\rm 3a1}(p^2,4,2,3) \notag \\
&\quad +U_{\rm 5b1,sub}(p^2,2,3,4,6,7) \,,\\[.5ex]
U_{\rm 5b1,sub}(...) &=
 \int_0^\infty ds_1\int_0^\infty ds_2 \; 
 \frac{\Delta B_0(s_1,6,7)}{(s_1-m_4^2-i\varepsilon)^2}\;
 \Delta B_0(s_2,2,3)\notag \\
&\hspace{10em} \times [B_0^{(0)}(p^2,s_1,s_2) - B_0^{(0)}(p^2,4,s_2)]
\,, \displaybreak[0]\\[1ex]
U_{\rm 5b2}(p^2,2,3,4,6,7) &= \tfrac{\partial}{\partial m_2^2}U_{\rm 5b}(0,2,3,4,6,7) 
 +B_0(4,6,7)\, T^{(0)}_{\rm 3a1}(p^2,2,3,4) \notag \\
&\quad +U_{\rm 5b2,sub}(p^2,2,3,4,6,7) \,,\\[.5ex]
U_{\rm 5b2,sub}(...) &=
 \int_0^\infty ds_1\int_0^\infty ds_2 \; 
 \frac{\Delta B_0(s_1,6,7)}{s_1-m_4^2-i\varepsilon}\;
 \Delta B_{0,m_1}(s_2,2,3)\, B_0^{(0)}(p^2,s_1,s_2) 
\,, \displaybreak[0]\\[1ex]
U_{\rm 5c1}(p^2,1,2,3,6,7) &= \tfrac{\partial}{\partial m_7^2}U_{\rm 5c}(0,1,2,3,6,7) 
 +T_{\rm 3a1}(1,7,6,3) \, B_0^{(0)}(p^2,1,2) \notag \\
&\quad +U_{\rm 5c1,sub}(p^2,1,2,3,6,7) \,,\\[.5ex]
U_{\rm 5c1,sub}(...) &=
 \int_0^\infty ds_1\int_0^\infty ds_2 \; 
 \frac{\Delta B_{0,m_1}(s_1,7,s_2)}{s_1-m_1^2-i\varepsilon}\;
 \Delta B_0(s_2,3,6) \, B_0^{(0)}(p^2,s_1,2) \,,
\end{align}
\begin{align}
U_{\rm 6m1}(p^2,1,2,3,4,6,8) &= \tfrac{\partial}{\partial m_3^2}U_{\rm 6m}(0,1,2,3,4,6,8)
 +B_0(0,6,8) \, T^{(0)}_{\rm 4a3}(p^2,1,2,3,4) \notag \\
&\quad + U_{\rm 6m1,sub}(p^2,1,2,3,4,6,8)  \,,\\[.5ex]
U_{\rm 6m1,sub}(...) &= \tfrac{\partial}{\partial m_3^2}U_{\rm 6m,sub}(...) \,,
\displaybreak[0]\\[1ex]
U_{\rm 6m3}(p^2,1,2,3,4,6,8) &= 
  \tfrac{\partial}{\partial m_2^2}U_{\rm 6m}(0,1,2,3,4,6,8) \notag \\
&\quad +B_0(0,6,8) \, T^{(0)}_{\rm 4a2}(p^2,1,2,3,4)
 + T_{\rm 4a}(0,0,0,6,8) \, B_{0,m_1}^{(0)}(p^2,2,1) \notag \\
&\quad + U_{\rm 6m3,sub}(p^2,1,2,3,4,6,8)  \,,\\[.5ex]
U_{\rm 6m3,sub}(...) &= \tfrac{\partial}{\partial m_2^2}U_{\rm 6m,sub}(...) \,,
\displaybreak[0]\\[1ex]
U_{\rm 6n2}(p^2,1,2,3,6,7,8) &= 
 \tfrac{\partial}{\partial m_2^2}U_{\rm 6n}(0,1,2,3,6,7,8) 
 + T_{\rm 4a}(0,0,0,3,6) \, B_{0,m_1}^{(0)}(p^2,2,1) \notag \\
&\quad +B_0(0,3,6) \, 
 \begin{aligned}[t] &[B_{0,m_1}(p^2,2,1)\,B_0(p^2,7,8) \\
 &- B_{0,m_1}(0,2,1)\,B_0(0,7,8)] \end{aligned} \notag \\
&\quad + U_{\rm 6n2,sub}(p^2,1,2,3,6,7,8)  \,, \displaybreak[0] \\[.5ex]
U_{\rm 6n2,sub}(...) &= \tfrac{\partial}{\partial m_2^2}U_{\rm 6n,sub}(...) \,,
\intertext{where}
T_{\rm 4a2}(p^2,1,2,3,4) &= \frac{\partial}{\partial m_2^2}T_{\rm 4a}(p^2,1,2,3,4)
 \notag \\
= \frac{1}{m_1^2+m_2^2-p^2} \biggl\{ \biggl [
 &\frac{2m_3^2(m_1^2-m_3^2+m_4^2)}{\sqrt{\lambda(1,3,4)}} \begin{aligned}[t] \bigl[ 
  &T_{\rm 3a1}(p^2,3,2,4) + (D-3)T_{\rm 4a}(p^2,1,2,3,4) \\
  &- B_0(0,3,3) B_0(p^2,1,2) \bigr] \biggr ] + \biggl[3
  \leftrightarrow 4 \biggr ] \end{aligned} \notag \\
 &-T_{\rm 3a1}(p^2,2,3,4) + (2D-7) T_{\rm 4a}(p^2,1,2,3,4) \notag \\
 &- 2m_1^2
 T_{\rm 4a1}(p^2,1,2,3,4)
 \biggr\}\,, \\[1ex]
T_{\rm 4a3}(p^2,1,2,3,4) &= \frac{\partial}{\partial m_3^2}T_{\rm 4a}(p^2,1,2,3,4)
 \notag \\
= \frac{1}{\sqrt{\lambda(1,3,4)}}\Bigl\{ &(m_1^2-m_3^2-m_4^2) 
\begin{aligned}[t]
 \bigl[&B_0(0,3,3)B_0(p^2,1,2)- T_{\rm 3a1}(p^2,3,2,4) \\
 & - (D-3)T_{\rm 4a}(p^2,1,2,3,4)
 \bigr ] \end{aligned} \notag \\
&  + 2m_4^2 \bigl [ B_0(0,4,4)B_0(p^2,1,2) - T_{\rm 3a1}(p^2,4,3,2)
 \bigr ] \Bigr\}
\end{align}
can be obtained from integration-by-parts identities. As described in
section~\ref{sc:tvid}, $\tfrac{\partial}{\partial m_3^2}U_{\rm 6m,sub}(...)$,
$\tfrac{\partial}{\partial m_2^2}U_{\rm 6m,sub}(...)$ and
$\tfrac{\partial}{\partial m_2^2}U_{\rm 6n,sub}(...)$ are
evaluated by means of a numerical differentiation in TVID~2.

\medskip\noindent
$U_{\rm 6m2}$ and $U_{\rm 6n1}$ are UV finite.
\begin{align}
U_{\rm 6m2}(p^2,1,2,3,4,6,8) &= -\int ds \; \Delta B_{0,m_1}(s,6,8)\,
 T_{\rm 5a}(p^2,1,2,3,4,s)\,, \\
U_{\rm 6n1}(p^2,1,2,3,6,7,8) &= -\int ds \; \Delta B_{0,m_1}(s,3,6)\,
 T_{\rm 5a}(p^2,1,2,s,7,8)\,.
\end{align}

%%%%%%%%%%%%%%%%%%%%%%%%%%%%%%%%%%%%%%%%%%%%%%%%%%%%%%%%%%%%%%

\section{TVID 2 manual}
\label{sc:manual}

\paragraph{Program name and version:} TVID, version 2.0 (August 2019).

\paragraph{System requirements:} {\sc Linux}-compatible platform; GNU C compiler {\sc gcc
4.4} or similar; {\sc Mathematica 10.x} \cite{mathematica}.

\paragraph{Copyright:} The TVID source code may be freely used and incorporated
into other projects, but the authors ask that always a reference to this
document and to Ref.~\cite{3lvac} be included.

\paragraph{External code elements:} TVID includes the Gauss-Kronrod routine
QAG from the {\sc Quadpack} library \cite{quadpack}, translated into C++, and
the C++ package {\sc doubledouble} for 30 digit floating point
arithmetic \cite{quad}.

It further requires the package {\sc LoopTools} \cite{looptools}, version 2.10
or higher, which must be installed by the
user separately.

\paragraph{Code availability:} The TVID source code is available for download at
\newline {\tt http://www.pitt.edu/\~{}afreitas/}.

\medskip

\subsection{Numerical part}

\noindent
The numerical part of TVID is programmed in C and evaluates the finite remainder
functions defined in appendix~\ref{sc:div}. It is called with the command
\begin{quote}
{\tt ucall} {\it infile} {\it outfile}
\end{quote}
where {\it infile} is the name of the input file, and {\it outfile} is the name
of the file where the results shall be placed. By default, {\tt ucall} is
located in the subdirectory {\tt ccode}. 

{\it infile} may contain a list
of lines, separated by line breaks, where each line has the form
\begin{quote}
{\it fname} \ {\it parA} \ {\it parB} \ \dots
\end{quote}
Here {\it fname} is the name of the function to be evaluated, see
Tabs.~\ref{tab:fnames1} and \ref{tab:fnames2}, and 
{\it parA}, {\it parB}, etc.\ are the numerical
momentum and mass parameters supplied. For example,
\begin{quote}
\tt U4 \ 1 \ 2 \ 3 \ 4 \\
\tt U5a \ 20 \ 1 \ 1 \ 1.5 \ 2 \ 2
\end{quote}
asks for the evaluation of $U_{4,\rm sub}(1,2,3,4)$ and of
$U_{\rm 5a,sub}(20,1,1,1.5,2,2)$. 
When $\tt ucall$ is completed, it fills {\it outfile} with a list of the
numerical results, again separated by line breaks. For instance, the example
above will return
\begin{quote}
\tt 
-5.555128856244808e1 0.0 \\
0.306188821751692 -6.207131465925367
\end{quote}
Here the first and second number in each row are the real and imaginary part of
the result, respectively.

%%%%%%%%%%%%%%%%%%%%%%%%%%%%%%%%%%%%%%%%%%%%%%%%%%%%%%%%%%%%%%
\begin{table}[tb]
\begin{center}
\begin{tabular}{|l|l|l|l|}
\hline
& & Symbol {\it fname} & Symbol used \\
Function & Input parameters & used by numerical & by algebraic \\
& & code {\tt ucall} & {\sc Mathematica} code \\
\hline
$U_{4,\rm sub}$ & $m_1,m_2,m_3,m_4$ & \tt \ U4 & \tt \ U4sub \\
$U_{4,\rm sub,0}$ & $m_2,m_3,m_4$ & \tt \ U40 & \tt \ U4sub0 \\
$U_{5,\rm sub}$ & $m_1,m_2,m_3,m_4,m_5$ & \tt \ U5 & \tt \ M21121 \\
$U_{5,\rm sub,0}$ & $m_3,m_4,m_5$ & \tt \ U50 & \tt \ M1p1121 \\
$U_{6,\rm sub}$ & $m_1,m_2,m_3,m_4,m_5,m_6$ & \tt \ U6 & \tt \ U6sub \\
\hline
\end{tabular}
\end{center}
\vspace{-2.5ex}
\mycaption{Symbols for basic finite remainder functions used in numerical and
algebraic parts of TVID 1 \cite{Bauberger:2017nct}.
\label{tab:fnames1}}
\end{table}
%%%%%%%%%%%%%%%%%%%%%%%%%%%%%%%%%%%%%%%%%%%%%%%%%%%%%%%%%%%%%%
\begin{table}[p]
\begin{center}\
\begin{tabular}{|l|l|l|l|}
\hline
& & Symbol {\it fname} & Symbol used \\
Function & Input parameters & used by numerical & by algebraic \\
& & code {\tt ucall} & {\sc Mathematica} code \\
\hline
$T_{\rm 3a,sub}$ & $p^2,m_2,m_3,m_4$ & \tt \ T3a & \tt T3asub \\
$T_{\rm 3a1,sub}$ & $p^2,m_2,m_3,m_4$ & \tt \ T3a1 & \tt T3a1sub \\
$T_{\rm 4a,sub}$ & $p^2,m_1,m_2,m_3,m_4$ & \tt \ T3a & \tt T3asub \\
$T_{\rm 4a1,sub}$ & $p^2,m_1,m_2,m_3,m_4$ & \tt \ T3a1 & \tt T3a1sub \\
$T_{\rm 5a}$ & $p^2,m_1,m_2,m_3,m_4,m_5$ & \tt \ T5a & \tt T5a \\
\hline \hline
$U_{\rm 4a,sub}$ & $p^2,m_1,m_3,m_6,m_8$ & \tt \ U4a & \tt U4asub \\
$U_{\rm 5a,sub}$ & $p^2,m_1,m_3,m_5,m_6,m_7$ & \tt \ U5a & \tt U5asub \\
$U_{\rm 5b,sub}$ & $p^2,m_2,m_3,m_4,m_6,m_7$ & \tt \ U5b & \tt U5bsub \\
$U_{\rm 5c,sub}$ & $p^2,m_1,m_2,m_3,m_6,m_7$ & \tt \ U5c & \tt U5csub \\
$U_{\rm 6a,sub}$ & $p^2,m_1,m_3,m_4,m_5,m_6,m_7$ & \tt \ U6a & \tt U6asub \\
$U_{\rm 6b,sub}$ & $p^2,m_1,m_3,m_4,m_5,m_6,m_8$ & \tt \ U6b & \tt U6bsub \\
$U_{\rm 6c,sub}$ & $p^2,m_1,m_2,m_3,m_4,m_6,m_7$ & \tt \ U6c & \tt U6csub \\
$U_{\rm 6m,sub}$ & $p^2,m_1,m_2,m_3,m_4,m_6,m_8$ & \tt \ U6m & \tt U6msub \\
$U_{\rm 6n,sub}$ & $p^2,m_1,m_2,m_3,m_6,m_7,m_8$ & \tt \ U6n & \tt U6nsub \\
$U_{\rm 7m,sub}$ & $p^2,m_1,m_2,m_3,m_4,m_5,m_6,m_8$ & \tt \ U7m & \tt U7msub \\
$U_{\rm 7a}$ & $p^2,m_1,m_2,m_3,m_4,m_6,m_7,m_8$ & \tt \ U7a & \tt U7a \\
$U_{\rm 8a}$ & $p^2,m_1,m_2,m_3,m_4,m_5,m_6,m_7,m_8$ & \tt \ U8a & \tt U8a \\
\hline \hline
$U_{\rm 4a1,sub}$ & $p^2,m_1,m_3,m_6,m_8$ & \tt \ U4a1 & \tt U4a1sub \\
$U_{\rm 4a2,sub}$ & $p^2,m_1,m_3,m_6,m_8$ & \tt \ U4a2 & \tt U4a2sub \\
$U_{\rm 4a3,sub}$ & $p^2,m_1,m_3,m_6,m_8$ & \tt \ U4a3 & \tt U4a3sub \\
$U_{\rm 5a1,sub}$ & $p^2,m_1,m_3,m_5,m_6,m_7$ & \tt \ U5a1 & \tt U5a1sub \\
$U_{\rm 5a2,sub}$ & $p^2,m_1,m_3,m_5,m_6,m_7$ & \tt \ U5a2 & \tt U5a2sub \\
$U_{\rm 5b1,sub}$ & $p^2,m_2,m_3,m_4,m_6,m_7$ & \tt \ U5b1 & \tt U5b1sub \\
$U_{\rm 5b2,sub}$ & $p^2,m_2,m_3,m_4,m_6,m_7$ & \tt \ U5b2 & \tt U5b2sub \\
$U_{\rm 5c1,sub}$ & $p^2,m_1,m_2,m_3,m_6,m_7$ & \tt \ U5c1 & \tt U5c1sub \\
$U_{\rm 6m1,sub}$ & $p^2,m_1,m_2,m_3,m_4,m_6,m_8$ & \tt \ U6m1 & \tt U6m1sub \\
$U_{\rm 6m2}$ & $p^2,m_1,m_2,m_3,m_4,m_6,m_8$ & \tt \ U6m2 & \tt U6m2 \\
$U_{\rm 6m3,sub}$ & $p^2,m_1,m_2,m_3,m_4,m_6,m_8$ & \tt \ U6m3 & \tt U6m3sub \\
$U_{\rm 6n1}$ & $p^2,m_1,m_2,m_3,m_6,m_7,m_8$ & \tt \ U6n1 & \tt U6n1 \\
$U_{\rm 6n2,sub}$ & $p^2,m_1,m_2,m_3,m_6,m_7,m_8$ & \tt \ U6n2 & \tt U6n2sub \\
$U_{\rm 7a1}$ & $p^2,m_1,m_2,m_3,m_4,m_6,m_7,m_8$ & \tt \ U7a1 & \tt U7a1 \\
$U_{\rm 7a2}$ & $p^2,m_1,m_2,m_3,m_4,m_6,m_7,m_8$ & \tt \ U7a2 & \tt U7a2 \\
\hline
\end{tabular}
\end{center}
\vspace{-2.5ex}
\mycaption{New symbols for basic finite remainder functions defined in TVID 2.
\label{tab:fnames2}}
\end{table}
%%%%%%%%%%%%%%%%%%%%%%%%%%%%%%%%%%%%%%%%%%%%%%%%%%%%%%%%%%%%%%

%\afterpage{\clearpage}

The option {\tt -e} allows the user to also receive information about the integration error:
\begin{quote}
{\tt ucall} {\it infile} {\it outfile} {\tt -e}
\end{quote}
In this case, a third number is added to each row in {\it outfile}, which
provides the integration error. For the example above, one obtains
\begin{quote}
\tt 
-5.555128856244808e1 0.0 \ 0.771470006356965e-10 \\
0.306188821751692 -6.207131465925367 \ 0.564457354626832e-11
\end{quote}

Internally, the numerical code uses the Gauss-Kronrod routine
QAG from the {\sc Quadpack} library \cite{quadpack} to evaluate the dispersion
integrals. This routine has been translated into C++ from the original FORTRAN
code, and amended to facilitate 30 digit floating point arithmetic from the
package {\sc doubledouble} \cite{quad}.

\subsection{Algebraic part}

\noindent
The algebraic part of TVID runs in {\sc Mathematica 10} \cite{mathematica} and
performs the separation of divergent and finite pieces of the master integrals.
The program is loaded in {\sc Mathematica} with
\begin{quote}
\tt << mcode/i3.m
\end{quote}
The two main user functions are {\tt PrepInt} and {\tt UCall} (and the variant
{\tt UCallE} of the latter).

\medskip\noindent 
{\tt PrepInt} takes as input any master integral in Figs.~\ref{fig:diag1},
\ref{fig:diag2} and \ref{fig:2lmaster}, 
as well as the three-loop vacuum master integrals $U_4$, $U_5$ or $U_6$, or a
linear combination thereof. It returns a series expansion in $\epsilon$, whose
coefficients contain the finite remainder functions listed in
Tabs.~\ref{tab:fnames1} and~\ref{tab:fnames2}. For example
\begin{quote}
\tt\small \rule{0mm}{0mm}\\[-2ex]
In[2]:=~PrepInt[SetPrecision[U4a[40,~1.1,~1.3,~1.6,~1.8],~30]];\\
\\
In[3]:=~N[\%]\\
\\
\rule{0mm}{0mm}~~~~~~~~~~~~~~~~~~~~~~~~~~~~4.15667~~~8.1407~~~11.2135\\
Out[3]=~(-734344.~+~0.~I)~+~-------~-~------~+~-------~+~\\
\rule{0mm}{0mm}~~~~~~~~~~~~~~~~~~~~~~~~~~~~~~~~~3~~~~~~~~2~~~~~\$eps\\
\rule{0mm}{0mm}~~~~~~~~~~~~~~~~~~~~~~~~~~~~~\$eps~~~~~\$eps\\
~\\
>~~~~U4asub[40.,~1.1,~1.3,~1.6,~1.8]~-~9767.45~U4sub[1.1,~1.3,~1.6,~1.8]~-~\\
~\\
>~~~~9844.68~U4sub[1.3,~1.1,~1.6,~1.8]~-~9844.53~U4sub[1.6,~1.1,~1.3,~1.8]~-~\\
~\\
>~~~~9767.1~U4sub[1.8,~1.1,~1.3,~1.6]\\[-2ex]
\end{quote}
\label{out3}
Here the directive {\tt SetPrecision} has been used to mitigate numerical
rounding errors within {\sc Mathematica}. {\tt PrepInt} can also be called with
symbols for the momentum and mass parameters, $e.\,g.$ {\tt
PrepInt[U4a[ps,m1s,m3s,m6s,m8s]]}, although this can lead to fairly large
expressions.

{\tt UCall} invokes the numerical code {\tt ucall} (see previous subsection)
to evaluate the finite
remainder functions in the output of {\tt PrepInt}. For the example above this
leads to
\begin{quote}
\tt\small \rule{0mm}{0mm}\\[-2ex]
In[4]:=~UCall[\%]\\
\\
\rule{0mm}{0mm}~~~~~~~~~~~~~~~~~~~~~~~~~~~~~~~~~4.15667~~~8.1407~~~11.2135\\
Out[4]=~(-149.694~+~9.60991~I)~+~-------~-~------~+~-------\\
\rule{0mm}{0mm}~~~~~~~~~~~~~~~~~~~~~~~~~~~~~~~~~~~~~~3~~~~~~~~2~~~~~\$eps\\
\rule{0mm}{0mm}~~~~~~~~~~~~~~~~~~~~~~~~~~~~~~~~~~\$eps~~~~~\$eps\\[-2ex]
\end{quote}
Technically, the executable {\tt ucall} is called through an external operating
system command, using the {\sc Mathematica} function {\tt Run}. The 
function {\tt UCall} looks for the executable {\tt ucall} in the subdirectory
{\tt ccode} of the TVID installation. If the user places {\tt ucall} in a
different directory, the variable {\tt \$Directory} in {\tt mcode/i3.m} must be
adjusted. For
passing input and output to and from the executable, {\tt UCall} uses the
filenames specified in the variables {\tt \$FileIn} and {\tt \$FileOut},
respectively. In most cases, the user will not need to change any of these
global variables.

The variant {\tt UCallE} also returns integration errors for the various
numerical master functions (by using the option {\tt -e} when calling {\tt
ucall}). For the example above this yields
\begin{quote}
\tt\small \rule{0mm}{0mm}\\[-2ex]
In[4]:=~UCallE[\%]\\
\\
\rule{0mm}{0mm}~~~~~~~~~~~~~~~~~~~~~~~~~~~~~~~~~4.15667~~~8.1407~~~11.2135\\
Out[4]=~(-734287.~+~9.60991~I)~+~-------~-~------~+~-------~-~\\
\rule{0mm}{0mm}~~~~~~~~~~~~~~~~~~~~~~~~~~~~~~~~~~~~~~3~~~~~~~~2~~~~~\$eps\\
\rule{0mm}{0mm}~~~~~~~~~~~~~~~~~~~~~~~~~~~~~~~~~~\$eps~~~~~\$eps\\
~\\
\rule{0mm}{0mm}~~~~~~~~~~~~~~~~~~~~~~~~~~~~~~~~~~~-11\\
>~~~~9767.45~(-22.9769~+~2.58504~10~~~~pm[1])~-~\\
~\\
\rule{0mm}{0mm}~~~~~~~~~~~~~~~~~~~~~~~~~~~~~~~~~~~-11\\
>~~~~9844.68~(-20.0775~+~2.30287~10~~~~pm[2])~-~\\
~\\
\rule{0mm}{0mm}~~~~~~~~~~~~~~~~~~~~~~~~~~~~~~~~~~~-11\\
>~~~~9844.53~(-16.7633~+~2.00197~10~~~~pm[3])~-~\\
~\\
\rule{0mm}{0mm}~~~~~~~~~~~~~~~~~~~~~~~~~~~~~~~~~~-11~~~~~~~~~~~~~~~~~~~~-9\\
>~~~~9767.1~(-15.0534~+~2.55672~10~~~~pm[4])~+~3.71208~10~~~pm[5]\\[-2ex]
\end{quote}
In this output, the numbers in front of $\tt pm[{\it n}]$ denote the errors of
the five finite remainder functions that are visible in the output $\tt Out[3]$
on page \pageref{out3}.

Examples for the use of TVID 2 can be found in the directory {\tt mcode}. {\tt
examples\_vaccum.m} demonstrates the use of 3-loop vacuum integrals, whereas {\tt
examples\_self.m} reproduces the numbers in Tabs.~\ref{tab:comp1} and~\ref{tab:comp2}.

\subsection{Installation}

TVID 2 comes in a compressed tar archive. After saving it in the desired
directory, it can be unpacked with the command
\begin{quote}
\tt tar xzf tvid.tgz
\end{quote}
The program contains the following subdirectory structure:
\\[1ex]
\begin{tabular}{ll}
{\tt ccode} & the C/C++ files for the numerical part of TVID 2;\\[.5ex]
{\tt ccode/doubledouble} & the {\sc doubledouble} for 30 digit floating point
arithmetic \cite{quad};\\[.5ex]
{\tt mcode} & the {\sc Mathematica} files for the algebraic part of TVID 2,\\\
& as well as examples.
\end{tabular}
\\[1ex]
Before compiling, the user must ensure that {\sc LoopTools} version 2.10 or
higher is installed on the system. It may be necessary to adjust the {\sc
LoopTools} path in {\tt ccode/makefile}.

\medskip\noindent
To compile the numerical C part of TVID, execute the commands
\begin{quote}
\tt cd ccode \\
\tt make
\end{quote}
The make file provided has been tested on Scientific Linux 6. It makes use of
the {\tt fcc} script included in {\sc LoopTools}, which should help to
facilitate compilation on a range of UNIX-type operating systems.  The authors
cannot guarantee that the installation process is successful on any operating
system, but they appreciate any helpful suggestions, comments and bug reports.

%%%%%%%%%%%%%%%%%%%%%%%%%%%%%%%%%%%%%%%%%%%%%%%%%%%%%%%%%%%%%%

\end{document}